\documentclass[times,twocolumn,final]{elsarticle}
\usepackage{algorithm,algorithmic}
\usepackage{framed,multirow}
\usepackage{amsmath}
\usepackage{amssymb}
\usepackage{latexsym}
\usepackage{url}
\usepackage{xcolor}
\usepackage{hyperref}
\usepackage{booktabs}
\usepackage{color}
\usepackage{geometry}
\graphicspath{{Figs/}}

\def\I{{\bf I}}

\def\L{{\bf L}}

\def\f{{\bf f}}

\def\m{{\bf m}}
\def\n{{\bf n}}

\def\u{{\bf u}}
\def\v{{\bf v}}

\def\x{{\bf x}}
\def\y{{\bf y}}
\def\z{{\bf z}}

\def\0{{\bf 0}}
\def\1{{\bf 1}}

\def\etal{{\em et al.}}
\def\eg{{\em e.g.}}
\def\ie{{\em i.e.}}

\def\etal{{\em et al.\/}\,}

\definecolor{newcolor}{rgb}{.8,.349,.1}

\begin{document}
%\verso{L.~Zhang \textit{et~al.}}

\begin{frontmatter} 
\title{Iterative Learning for Joint Image Denoising and Motion Artifact Correction of 3D Brain MRI}

\author{Lintao Zhang, Mengqi Wu, Lihong Wang, David C. Steffens, Guy G. Potter, Mingxia Liu\corref{cor1}}

\cortext[cor1]{Corresponding author: M.~Liu (mingxia\_liu@med.unc.edu).}

\begin{abstract}
Image noise and motion artifacts greatly affect the quality of brain magnetic resonance imaging (MRI) and negatively influence downstream medical image analysis. 
Previous studies often focus on 2D methods that process each volumetric MR image slice-by-slice, thus losing important 3D anatomical information. 
Additionally, these studies generally treat image denoising and artifact correction as two standalone tasks, without considering their potential relationship, especially on low-quality images where severe noise and motion artifacts occur simultaneously. 
To address these issues, we propose a Joint image Denoising and motion Artifact Correction (JDAC) framework via iterative learning to handle noisy MRIs with motion artifacts, consisting of an \emph{adaptive denoising model} and an \emph{anti-artifact model}.   
In the adaptive denoising model, we first design a novel \emph{noise level estimation strategy}, and then adaptively reduce the noise through a U-Net backbone with feature normalization conditioning on the estimated noise variance. 
The anti-artifact model employs another U-Net for eliminating motion artifacts, incorporating a novel \emph{gradient-based loss} function designed to maintain the integrity of brain anatomy during the motion correction process. 
These two models are iteratively employed for joint image denoising and artifact correction through an iterative learning framework. 
An early stopping strategy depending on noise level estimation is applied to accelerate the iteration process. 
The denoising model is trained with {\color{black}9,544} T1-weighted MRIs with manually added Gaussian noise as supervision. 
The anti-artifact model is trained on {\color{black}552} T1-weighted MRIs with motion artifacts and paired motion-free images. 
We validate the proposed method on a public dataset and a clinical study that involves MRIs distorted by motion and noise. 
Experimental results suggest the effectiveness of JDAC in both tasks of denoising and motion artifact correction, compared with several state-of-the-art methods. 
\end{abstract}
\end{frontmatter}

\section{Introduction}
\label{S1}
{\color{black}
{M}edical image denoising and motion artifact correction are key to MRI processing~\citep{sagheer2020review, kaur2023complete}, which directly determines the processed image quality and affects downstream analysis~\citep{budrys2018artifacts, kaur2023complete}.
Medical images typically have three-dimensional (3D) volumetric data and low contrast, making them susceptible to noise and artifacts.
However, most of previous denoising and anti-artifact methods are 2D-based and applicable to natural images. 
When applied to 3D medical images like brain MRI, those methods have to be performed slice-by-slice, resulting in the loss of important 3D anatomical information.
For instance, some studies~\citep{adame2023fondue,al2022stacked} try to denoise or correct motion artifacts of 3D MRI using 2D-based models on different imaging planes.
Many studies on motion artifact correction~\citep{liu2020motion,  al2022stacked, sommer2020correction} have attempted to synthesize motion-corrupted 2D MRI slices to train anti-artifact models.
Xiang et al.~\citep{xiang2022ddm} demonstrate that when brain diffusion MRI is processed slice-by-slice on the axial plane using 2D models, it causes obvious discontinuities (gaps or breaks in image quality) on the other two planes (coronal and sagittal) compared to results of 3D models.
Several recent studies have tried different strategies to make use of 3D structural information in medical images.
Some traditional methods are re-implemented to denoise 3D volumetric data, such as BM4D~\citep{makinen2020collaborative, makinen2022ring}. 
Some studies rely on 3D convolutional neural networks (CNN) to directly handle the task of denoising~\citep{hou2022truncated, xiang2022ddm} or anti-artifact~\citep{duffy2021retrospective} of 3D medical images.
Besides, the above-mentioned denoising or anti-artifact models are usually designed without considering their potential relationship. % on each other.
Many established pipelines for medical image processing~\citep{backhausen2016quality, cai2021prequal, maximov2019towards} also perform these two tasks separately.
This may lead to a sub-optimal MRI processing result when severe noise and motion artifacts occur simultaneously. 

On the other hand, the alternating direction method of multiplier (ADMM) method~\citep{boyd2011distributed} has attracted more attention recently, due to its good performance in reducing MRI reconstruction noise and artifacts {\color{black}in an iterative manner}.
For example, Chan et al.~\citep{chan2016plug} use the Plug-and-play ADMM (PnP-ADMM) algorithm for iterative image restoration and prove the algorithm convergence under a bounded denoiser assumption. 
Hou et al.~\citep{hou2022truncated} demonstrate that the PnP-ADMM algorithm can be used to gradually remove MRI reconstruction noise caused by sparse sampling in the $K$-space. 
Currently, few studies have attempted to use iterative learning strategies to perform denoising and motion artifact correction tasks to progressively improve MRI image quality.

\begin{figure*}[!t]
\setlength{\belowdisplayskip}{-0pt}
\setlength{\abovedisplayskip}{-0pt}
\setlength{\abovecaptionskip}{-0pt}
\setlength{\belowcaptionskip}{-0pt}
\centering
\includegraphics[width=0.98\textwidth]{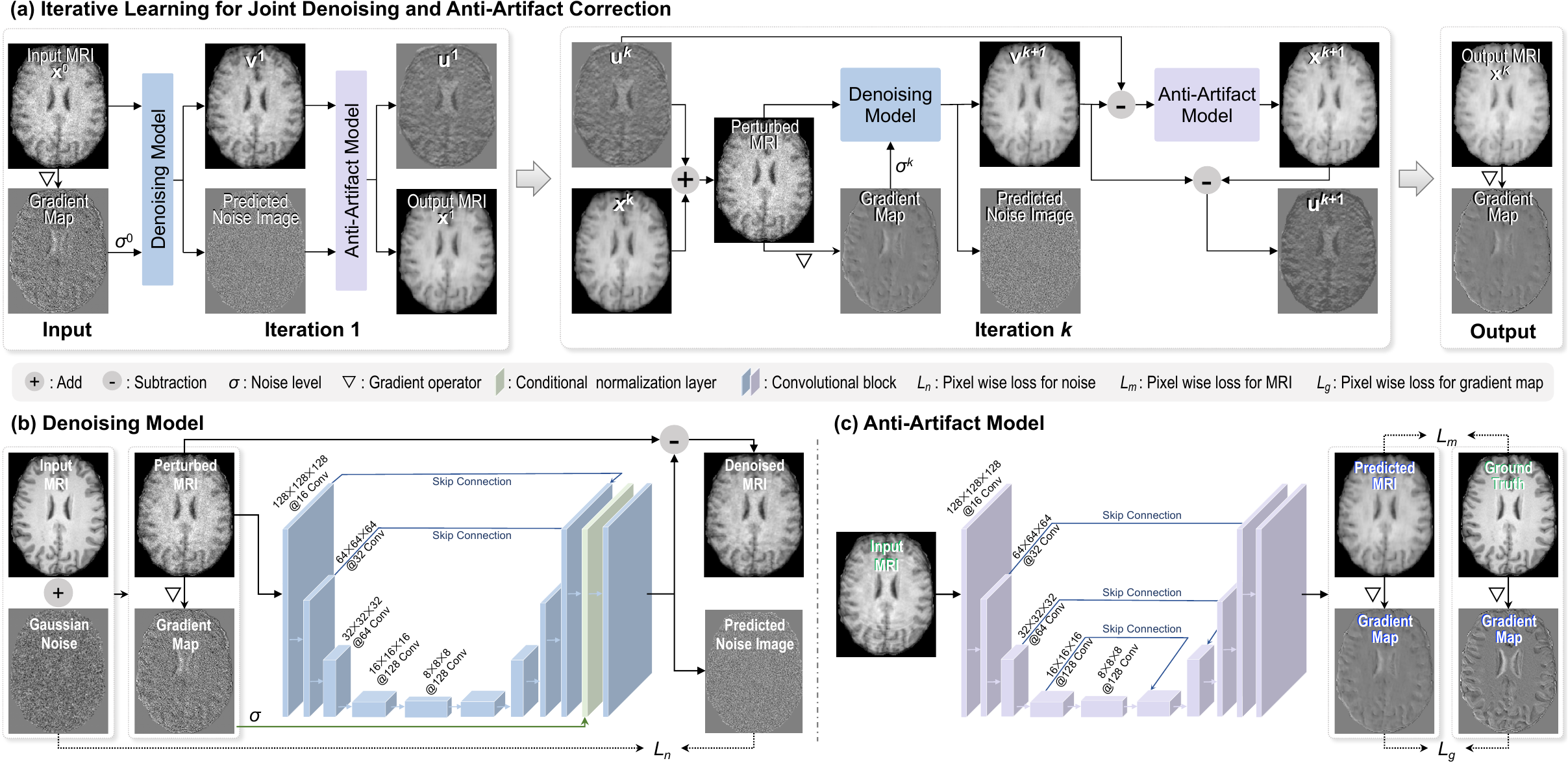}
\caption{Illustration of the proposed iterative learning framework (\ie, panel (a)) of joint image denoising and motion artifact correction (JDAC) for structural MRI data processing. 
The JDAC consists of an \emph{adaptive denoising model} (\ie, panel (b)) and an \emph{anti-artifact model} (\ie, panel (c)) that iteratively reduces the MRI image noise and motion artifacts. 
The denoising model can adaptively denoise the noisy MRI based on estimated noise levels.
The anti-artifact model is trained with motion-free MRIs as ground truth and constrained by a new gradient-based loss function for brain structure preserving.}
\label{figFramework}
\end{figure*}

To this end, we propose an iterative framework to jointly perform image denoising and motion artifact correction (JDAC) for T1-weighted brain MRIs. 
As shown in Fig.~\ref{figFramework} (a), the JDAC incorporates an \emph{adaptive denoising model} and an \emph{anti-artifact model} in an iterative learning manner. 
In the adaptive denoising model (see Fig.~\ref{figFramework} (b)), we first propose a new method to estimate image noise levels using the variance of the image gradient map and validate the method through statistical analysis. 
An adaptive denoising model with a U-Net architecture is then developed to denoise an MR image based on the estimated noise level. 
The anti-artifact model employs another U-Net for eliminating motion artifacts, incorporating a novel gradient-based loss function designed to maintain the integrity of brain anatomy during the motion correction process, as shown in Fig.~\ref{figFramework} (c).
Extensive experiments and ablation studies are performed to validate the effectiveness of JDAC.
The results show that jointly performing denoising and motion artifact correction tasks via the iterative learning strategy can progressively improve MR image quality, especially for motion-affected MRIs with severe noise.} 
The source code has been released to the public via GitHub\footnote{https://github.com/goodaycoder/JDAC}.

The main contributions of this work are summarized below.
%\vspace{-2mm}
\begin{itemize}
\vspace{-2mm}
    \item 
    A joint image denoising and motion artifact correction (JDAC) framework is developed to iteratively handle noisy MRI with motion artifacts. 
    By implicitly exploring underlying relationships between denoising and artifact correction tasks, the proposed JDAC is expected to progressively improve image quality.

    \vspace{-2mm}
    \item
    We design an adaptive denoising model, where a new noise level estimation strategy is designed that uses the variance of gradient maps to quantitatively estimate noise levels. 
    The noise level estimation result can be used as the condition of the denoiser and a threshold in the early stopping strategy for our iterative learning framework. 
    
    \vspace{-2mm}
    \item We further introduce an innovative gradient-based loss function in the anti-artifact model, aiming at retaining brain anatomy details throughout the motion correction procedure. This helps ensure that the model does not distort the original brain structures in 3D MR images.

    \vspace{-2mm}
    \item Extensive experiments have been performed on both tasks of adaptive denoising and joint denoising and motion artifact on two public datasets and a real motion-affected MRI dataset in a clinical study. 
    Quantitative and qualitative results suggest the effectiveness of our JDAC. 
\end{itemize}

The remainder of this paper is organized as follows. 
Section~\ref{S2} reviews the most relevant studies. 
Section~\ref{S3} introduces the proposed framework in detail. 
Section~\ref{S4} introduces experimental setup, competing methods, and experimental results. 
Section~\ref{S5} analyzes the influences of several key components of JDAC and discusses the limitations of this work and future research directions. 
This paper is concluded in Section~\ref{S6}.

%%%%%% -- New Section -- %%%%%%
%%%%%%%%%%%%%%%%%%%%%%%%%%%%%%%%%%%%%%%%%%%%%%%%%%%%%%%%%%%%%%%
\section{Related Work}
\label{S2}
\subsection{MR Image Denoising}%  of Medical Images}
Much denoising research has been carried out in the field of natural image processing, and classic methods such as BM3D~\citep{dabov2007image} have been applied to medical image denoising.
Most studies on image denoising focus on 2D-based methods that have to be performed in a slice-by-slice way when analyzing 3D medical images like MRI scans, CT scans,  X-rays, and ultrasound imaging~\citep{kaur2023complete}. 
Some studies have extended classical 2D methods to 3D versions such as BM4D~\citep{makinen2020collaborative, makinen2022ring} to facilitate 3D image denoising. 

Recently, many deep learning models have been created for image denoising. 
For example, Lehtinen~\etal~\citep{lehtinen2018noise2noise} reveal that it is feasible to train a deep learning model for image denoising using solely corrupted samples, achieving and occasionally surpassing the performance of training with clean data. 
Many studies show an effective way to train a deep denoising model by manually adding zero-mean white noise~\citep{lehtinen2018noise2noise, kim2021noise2score}. 
Ho~\etal~\citep{ho2020denoising} propose to use a denoising diffusion probabilistic model (DDPM), demonstrating that image noise can be reduced multiple times gradually. 
The most recent research using diffusion models~\citep{xiang2022ddm, gong2023pet,zhu2023denoising} has shown better denoising performance and is getting more attention in medical image denoising and restoration.
For example, Zhu~\etal~\citep{zhu2023denoising} try to combine a plug-and-play method and a diffusion model to make use of the generative ability of diffusion models for better image deblurring. 
Xiang~\etal~\citep{xiang2022ddm} propose a framework that integrates statistic-based denoising theory and diffusion Markov chain to perform self-supervised denoising of diffusion MRI (dMRI). 
Moreover, researchers usually rely on the plug-and-play alternating direction method of multiplier (PnP-ADMM)~\citep{boyd2011distributed}  for MRI reconstruction, where a pre-trained denoiser is used to iteratively reduce the reconstruction noise caused by fast sparse sampling in $k$-space~\citep{dong2018denoising, ryu2019plug}.
For instance, Hou~\etal~\citep{hou2022truncated} propose a PnP-ADMM that truncates the predicted residual of the denoiser to iteratively reduce the reconstruction noise. 
All the studies show that reducing the medical image noise iteratively may be a more effective way. 

Inspired by these studies that use progressive denoising, this paper considers an iterative learning strategy to perform denoising of brain structural MRIs. 
Additionally, even though DDPM-based models~\citep{xiang2022ddm, gong2023pet,zhu2023denoising} and ADMM-based methods~\citep{dong2018denoising, ryu2019plug, hou2022truncated} focus on progressively removing noise, they cannot explicitly estimate the noise level conveyed in input images. 
Intuitively, it is interesting to utilize such prior knowledge (\ie, noise level) for adaptive denoising to boost learning performance. 

%\subsection{Motion Artifact Correction with Small MRI Data}
\subsection{Motion Artifact Correction of MRI}
For learning-based motion artifact correction studies, the issues of over-smoothing, decreasing image contrast, and distortion of small anatomic structures are very important during the motion artifact reduction process~\citep{chang2023deep}. 
In previous studies, researchers proposed many neural network architectures to learn simulated artifact errors precisely.
For example, Liu~\etal~\citep{liu2020motion} propose a 2D deep CNN with multi-resolution blocks to predict a residual artifact image of the simulated motion artifacts, thus trying to avoid image contrast loss. 
Al-Masni~\etal~\citep{al2022stacked} develop a retrospective 2D stacked UNet to learn the rigid motion artifacts in a coarse-to-fine manner with synthesized motion-corrupted images. 
Duffy~\etal~\citep{duffy2021retrospective} train a motion correction CNN with 3D simulated artifacts on motion-free MRI scans, improving the cortical surface reconstruction quality.
These models trained with simulated artifacts can accurately predict residual artifacts on simulated data and preserve brain structural details well, but their performance usually drops dramatically on real motion-affected data. 
Recently, N{\'a}rai~\etal~\citep{narai2022movement} publish a movement-related artefacts (MR-ART) dataset of structural brain MR images with ground-truth motion artifacts. 
As reported in a recent study~\citep{safari2023mri}, the performance of these anti-artifact models on the MR-ART dataset declines compared to the test results on simulated data reported in the above studies. 
One of the most likely reasons is that these studies only used regular loss functions for model training without explicit constraints on preserving brain anatomy. 
In this work, we propose a new gradient-based loss function to constrain the anti-artifact model for brain structure preservation. 

In addition, most previous studies intended for denoising or motion artifact correction ignore the potential relationship between these two tasks. 
Some studies~\citep{zhang2021plug} have demonstrated that the denoising model will not only remove noise but also remove some artifacts and image texture details. 
Classical medical image processing pipelines usually include one of the two tasks. 
For instance, Backhausen et al.~\citep{backhausen2016quality} propose a quality control workflow to rate motion artifacts as a complement to automated processing tools like FreeSurfer~\citep{fischl2012freesurfer}.
Cai et al.~\citep{cai2021prequal} try to combine the denoising and motion-induced artifact removal from different tools (\eg, FSL~\citep{jenkinson2012fsl}, MRTrix3~\citep{tournier2019mrtrix3}, and ANTs~\citep{avants2009advanced}) for integrated preprocessing of diffusion-weighted MR imaging (DWI).
Maximov et al.~\citep{maximov2019towards} evaluate a general pipeline including both noise correction and Gibbs artifact removal, demonstrating the pipeline can be optimized for DWI.
Inspired by these studies, we consider jointly performing denoising and motion artifact correction to handle noisy MRI with motion artifacts.
By implicitly exploring underlying relationships between these two tasks, our method is expected to further improve image quality.

%%%%%% -- New Section -- %%%%%%
\section{Methodology}
\label{S3}
\if false
In this section, we first introduce an iterative learning strategy for joint image denoising and artifact correction. 
Then, we present a new image noise level estimation method and an adaptive denoising model. 
Finally, the motion artifact reduction model is presented in detail. 
\fi 

%%%%%%%%%%%%%%%%%%%%%%%%%%% new section   new section  new section %%%%%%%%%%%%%%%%%%%%%%%%%%%%%%%%%%%
\subsection{Proposed Method}
In clinical applications, MRI quality is easily degraded from both imaging noise and motion artifacts.
While many state-of-the-art methods have been proposed recently for image denoising and anti-artifacts, most of the methods and processing pipelines reduce the noise and motion artifacts separately.
Ignoring the mutual interaction between noise and motion artifacts may lead to suboptimal post-processing results.
In this work, we propose an iterative learning strategy to perform joint image denoising and motion artifact correction (JDAC), thus implicitly exploring underlying relationships between denoising and artifact correction tasks. 
From our experimental observations, it appears that motion artifacts closely resemble certain structural textures in MRI. 
In contrast, additive noise tends to exhibit a relatively independent distribution.
In addition, motion artifact correction results are often greatly affected by denoising results, while it could be easier to model noise information than motion artifacts. 
Therefore, we propose to perform image denoising and motion artifact correction sequentially in each iteration. 
As shown in Fig.~\ref{figFramework}, the JDAC incorporates (1) an \emph{adaptive denoising model} for noise removal,  and (2) an \emph{anti-artifact model} for motion artifact reduction, both equipped with UNet-like architectures. % for 3D MRI feature learning. 
These two models are jointly utilized for image denoising and artifact correction through an iterative learning framework. 
%More details can be found in the following. 

\subsection{Proposed Iterative Learning Framework}

\subsubsection{Iterative Learning Strategy}

In this work, we propose a novel iterative learning strategy for joint denoising and motion artifact correction.  
%Denote $L \times W \times H$ as the size of a 3D MR image.  
Given an MRI of ideal quality as $\x \in \mathbb{R}^{L \times W \times H}$, the motion-affected noisy measurement $\y \in \mathbb{R}^{L \times W \times H}$ can be represented as
\begin{equation}
\label{eq_y}
\small
 \y = \mathcal{A}(\x) + \xi\\
\end{equation}
where $\mathcal{A}$ denotes a distortion function due to motion and $\xi$ is additive noise. 
Without specification, we consider Gaussian noise in this work. 
Mathematically, we formulate the problem of joint MRI denoising and motion artifact correction as: 
\begin{equation}
\label{eq_optim}
\small
 \hat{\x} = \mathop{argmin}\limits_{\x}|| \mathcal{A}(\x)-\y||_2^2 + \mathcal{D}(\x)
\end{equation}
where the first term is used for motion artifact correction and $\mathcal{D}(\x)$ is employed for denoising. 

The problem in equation Eq.~\eqref{eq_optim} can be solved by the alternating direction
method of multiplier (ADMM)~\citep{boyd2011distributed}. 
Through variable substitution, Eq.~\eqref{eq_optim} can be reformulated as a constrained problem as follows:
\begin{equation}
\label{eq_argmin}
\small
  \mathop{arg min}\limits_{\x,\v}||\mathcal{A}(\x)-\y||_2^2 + \mathcal{D}(\v), \, \mathop{s.t.} \x = \v
\end{equation}

The augmented Lagrangian function of Eq.~\eqref{eq_argmin} can be written as:
\begin{equation}
\label{eq_Lagrangian}
\small
\mathcal{L}(\x,\v,\u) = ||\mathcal{A}(\x)-\y||_2^2 + \mathcal{D}(\v) + \u^T (\x - \v) + \frac{\rho}{2} ||\x - \v||_2^2
\end{equation}
where $\u$ and $\rho$ are Lagrange multipliers. 
We can solve the following subproblems to get the saddle
point of $\mathcal{L}$, which yields the minimizer of Eq.~\eqref{eq_Lagrangian}:
\begin{equation}
\label{eq_minimizer_v}
\small
\v^{k+1} =\mathop{arg min}\limits_{\v} \mathcal{D}(\v) + \frac{\rho}{2} ||(\x^k+\u^k) - \v||_2^2
\end{equation}
\begin{equation}
\label{eq_minimizer_x}
\small
\x^{k+1} = \mathop{arg min}\limits_{x}||\mathcal{A}(x)-\y||_2^2 + \frac{\rho}{2} ||\x^k-(\v^{k+1}-\u^k)||_2^2
\end{equation}
\begin{equation}
\label{eq_minimizer_u}
\small
\u^{k+1} = \u^{k} + (\x^{k+1} - \v^{k+1})
\end{equation}

The subproblem defined in Eq.~\eqref{eq_minimizer_v} can be treated as a denoising
problem~\citep{chan2016plug}, rewritten as: % and rewrite as:
\begin{equation}
\label{eq_minimizer_v1}
\small
\v^{k+1} =\mathop{arg min}\limits_{\v} \mathcal{D}(\v) + \frac{1}{\sigma^2} || \v - \tilde{\v}^k||_2^2
\end{equation}
in which $\tilde{\v}^k = (\x^k+\u^k)$ denotes the noisy image degraded by the
Gaussian noise with standard deviation (std) $\sigma$. 
In this work, we propose a noisy estimation strategy to obtain this hyperparameter $\sigma^k$ at the $k$-th iteration. 

This problem can be solved by training a  denoising model $\f_{D}$ iteratively~\citep{chan2016plug,dong2018denoising,venkatakrishnan2013plug} through the following: 
\begin{equation}
\small
\label{eq_denoiser}
\v^{k+1} = \f_{D}(\tilde{\v}^{k}, \sigma^k) = \f_{D}(\x^k+\u^k, \sigma^k)
\end{equation}
where the $\u^k$ can be regarded as the predicted residual component that updates based on Eq.~\eqref{eq_minimizer_u} during the iteration.

The subproblem of motion artifact correction defined in Eq.~\eqref{eq_minimizer_x} can also be solved by iteratively learning a deep neural network $\f_{A}$, formulated as:
\begin{equation}
\label{eq_antiart}
\small
 \x^{k+1} = \f_{A}(\tilde \x^{k}) = \f_{A}(\v^{k+1}-\u^{k}) 
\end{equation}
where the $\tilde \x^{k} = (\v^{k+1}-\u^{k})$ denotes the $k$-th motion-affected MR image during iterations.

Inspired by~\citep{zhang2021plug,hou2022truncated}, we use two pretrained models for denoising and anti-artifact in this work.
Previous studies also demonstrate that denoising models, when applied with the ADMM iteration strategy, tend to over-smooth noise, artifacts, aliasing, and image details~\citep{zhang2021plug,hou2022truncated}.
Similarly, our experimental observations discover that MR images can also become over-smoothed when the anti-artifact model is applied multiple times (\eg, more than 5 times).
Intuitively, the texture features of motion artifacts are very similar to original texture features of an MR image, so multiple removals may lead to the over-smoothing problem. 
To mitigate this issue, we propose to update the $\x^{k}$ with a learning rate $\delta$, and the $\u^{k+1}$ in Eq.~\eqref{eq_minimizer_u} without accumulation. 
This can be expressed as: 
\begin{equation}
\label{eq_lr}
\small
 \x^{k} = \x^{k} \times (1-\delta) + \v^{k} \times \delta
\end{equation}
\begin{equation}
\label{eq_update_u}
\small
 \u^{k+1} = \x^{k+1}-\v^{k+1}
\end{equation}

Meanwhile, the estimated noise level of the corrected MR images (\ie, $\sigma^k$) in each iteration is used as an early stopping criterion. %to end the iterative process early. 
That is, the iteration ends when the noise level sigma is lower than a threshold $\Delta$, where the threshold is empirically determined using an average std value of gradient maps of clean MRIs.
For instance, we set $\Delta = 0.028$ in our work based on Fig.~\ref{fig_NoiseLevelEstiamtion}~(b).
With the early stopping strategy, we found our JDAC typically requires only
1 or 2 iterations to perform joint denoising and motion artifact correction. 
More discussion is given in Section~\ref{S_DiscussionIterative}. 
The detailed implementation of the proposed iterative learning algorithm is shown in Algorithm~\ref{alg_jdac}, with the flowchart illustrated in Fig.~\ref{figFramework}~(a). 
%%%%%%%%%%%%%%%%%%%%%%%%%%%%%%%%%%%%%%%%%%%%%%%
\begin{algorithm}
%\small
\footnotesize
    \caption{{\small{Proposed Iterative Learning Algorithm of JDAC}}}
    \label{alg_jdac}
    \begin{algorithmic}[1]
        \REQUIRE Denoising model $\f_{D}$, Anti-artifact model $\f_{A}$, Noise estimation $\f_{N}$, Input ${\y}$, Learning rate $\delta$, Max iterations $K$, Early-stop threshold $\Delta$
        \ENSURE Denoised and motion corrected MRI: $\hat{\x}$  
        \STATE Initialization: $\x^{0}=\v^{0}\leftarrow \y$,  $\u^{0} \leftarrow 0$
        \FOR {$k = 0$ to $K-1$}          
            %\STATE $\v_{old} = \v$, $\u_{old} = \u$
            \STATE $\x^{k} = \x^{k} \times (1-\delta) + \v^{k} \times \delta$
            \STATE $\tilde{\v}^{k+1} = \x^{k} + \u^{k}$
            \STATE $\sigma^{k}_{e} = \f_{N}(\tilde{\v}^{k+1})$
            \STATE $\v^{k+1} = \f_{D}(\tilde{\v}^{k+1}, \sigma^{k}_{e})$\\
            \STATE $\tilde \x^{k+1} = \v^{k+1}-\u^{k}$\\
            \STATE $\x^{k+1} = \f_{A}(\tilde \x^{k+1})$\\
            \STATE $\u^{k+1} = \x^{k+1}-\v^{k+1}$\\
%            \STATE $\hat{\x} = clip[\x^{k+1}, 0, 1]$ 
            %\STATE $\hat{\x} = \x^{k+1}$             
            %\STATE $\sigma_{e} = \f_{N}(\hat{\x})$
            \STATE $\sigma_{e} = \f_{N}(\x^{k+1})$
            \IF { $\sigma_{e} <\Delta$}
                \RETURN $\hat{\x}$
            \ENDIF
        \ENDFOR
    \end{algorithmic}
\end{algorithm}    
%%%%%%%%%%%%%%%%%%%%%%%%%%%%%%%%%%%%%%%%%%%%%%%

\subsubsection{Noise Estimation Strategy of Structural MRI}
\begin{figure}[!t]
\setlength{\belowdisplayskip}{0pt}
\setlength{\abovedisplayskip}{0pt}
\setlength{\abovecaptionskip}{0pt}
\setlength{\belowcaptionskip}{0pt}
\centering
\includegraphics[width=0.48\textwidth]{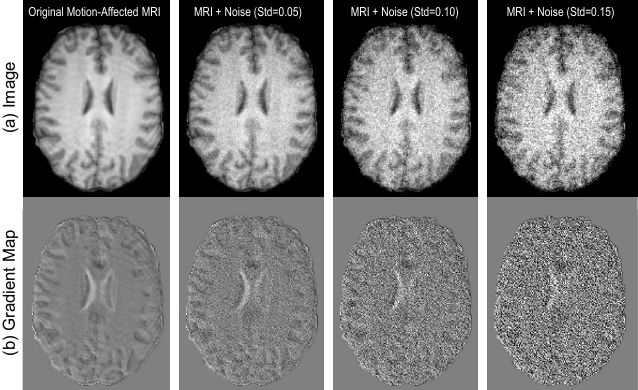}
\caption{Illustration of an original motion-affected MRI from MR-ART~\citep{narai2022movement} with different manually added Gaussian noise levels (top) and their corresponding gradient maps (bottom). The contrast of gradient maps is normalized for better visualization. Std: standard deviation.}
\label{fig_NoiseLevels}
\end{figure}

In image-denoising studies that manually add noise level onto a clean image as ground truth, researchers usually use noise variance such as standard deviation (std) to represent noise level in images~\cite{mohan2014survey}. 
For many denoising methods ~\citep{nichol2021improved,xiang2022ddm,kawar2022denoising}, the variance of the noise level is a key factor to denoising results, especially for most iterative frameworks that gradually reduce the noise (\eg, the most popular DDM-based models~\citep{nichol2021improved}). 
When given a real noisy image without any prior on the noise, it is challenging to explicitly estimate the noise level. 
In Fig.~\ref{fig_NoiseLevels}~(a), we manually add Gaussian noise with std of [0.05, 0.10, 0.15] to a high-quality clean MRI with motion artifacts (image intensity std of 0.306). 
The gradient maps of the original MRI and three perturbed images are shown in Fig.~\ref{fig_NoiseLevels}~(b), from which we can see that the \emph{std of gradient maps is sensitive to different noise levels}.  
Inspired by this finding, we propose to employ std of gradient maps of MRIs to estimate image noise level.

\begin{figure*}[!t]
\setlength{\belowdisplayskip}{0pt}
\setlength{\abovedisplayskip}{0pt}
\setlength{\abovecaptionskip}{0pt}
\setlength{\belowcaptionskip}{0pt}
\centering
\includegraphics[width=0.99\textwidth]{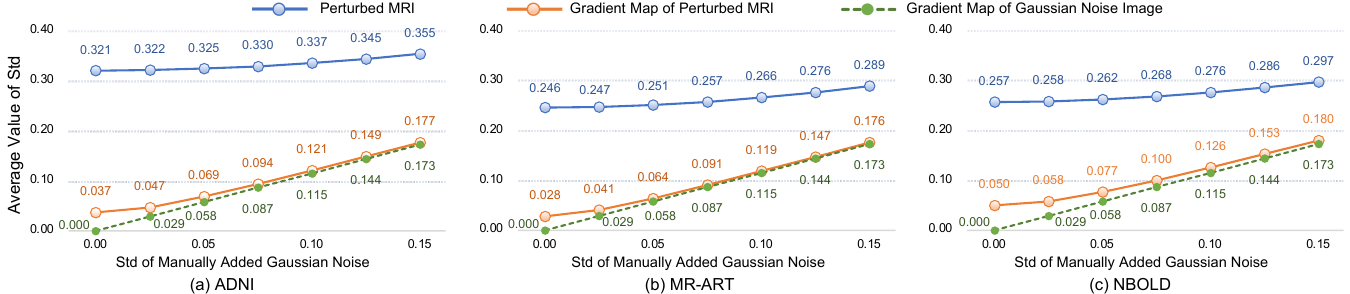}
\caption{Average standard deviation (std) values of perturbed MR images (\ie, MRI with manually added Gaussian noise), their corresponding gradient maps, and gradient maps of the added Gaussian noise. These averaged values are calculated based on 40 MRIs randomly selected from (a) ADNI~\citep{jack2008alzheimer}, (b) MR-ART~\citep{narai2022movement}, and (c) NBOLD~\citep{steffens2017negative}.
}
\label{fig_NoiseLevelEstiamtion}
\end{figure*}

To quantitatively assess the relationship between manually added noise levels and the std values of image gradient maps, we randomly sample 40 MRI scans from three different datasets, including (1) Alzheimer's Disease Neuroimaging Initiative (ADNI)~\citep{jack2008alzheimer}, (2) Movement-Related Artifacts (MR-ART) dataset~\citep{narai2022movement}, and (3) Neurobiology of Late-life Depression (NBOLD) study~\citep{steffens2017negative}. 
In Fig.~\ref{fig_NoiseLevelEstiamtion}, we report the average intensity std values of perturbed MRI (MRI+noise) and their gradient maps with the increase of noise levels. 
We also show the average std values of gradient maps of the added Gaussian noise (green dashed lines in Fig.~\ref{fig_NoiseLevelEstiamtion}). 
From Fig.~\ref{fig_NoiseLevelEstiamtion}, we can get several interesting observations.
(1) When there is no noise added, the average std values of the MRI gradient maps in three different datasets (\ie, 0.037, 0.028, and 0.050, respectively) are much lower than that of the original MR images (\ie, 0.321, 0.246, and 0.257, respectively).
(2) With the increase of noise levels, the std values of perturbed MRIs increase slightly (see blue lines), while the std values of their gradient maps tend to increase linearly (see orange lines).
(3) The gradient map std values of perturbed MRIs and manually added Gaussian noise are close to each other, especially when the noise std is larger than 0.025.  
Based on the above observations, we employ the std of image gradient maps to estimate noise levels in brain MRIs without any prior knowledge of image noise. 
On the other hand, Figure~\ref{fig_NoiseLevelEstiamtion} suggests that, even though the average std values of perturbed MRI scans in different databases vary greatly, those values of their gradient maps are very consistent at each noise level. 
This implies that the std of gradient maps can be used as a general metric to estimate noise levels. 

%%%%%%%%%%%%%%%%%%%%%%%%%%%%%%%%%%%%%%%%%%%%%%%
%%%%%%%%%%%%%%%%%%%%%%%%%%%%%%%%%%%%%%%%%%%%%%%%%%%%%%%%%%%%%%%%%%%%%%%%%%%%%%%%%%%%%%%%%%%%%%
%%%%%%%%%%%%%%%%%%%%%%%%%%%%%%%%%%%%%%%%%%%%%%%

Mathematically, we propose to estimate the noise level in an MR image $\x$ through the following:  
\begin{equation}
\label{eq_noiselevel}
\small
 \sigma_{e} = \f_{N}(\x) \approx \sqrt{Var(\bigtriangledown \x)}
\end{equation}
where the $\sigma_{e}$ denotes the estimated std of MRI image noise, $\bigtriangledown$ is the gradient operator. 
Instead of simply computing an approximation using the variance of the gradient map as in Eq.~\eqref{eq_noiselevel}, one can also use numerical fitting strategies to find the optimal  $\f_{N}$ given a set of data with ground-truth noise. 
In our JDAC framework, we use the std of image gradient maps as the noise level estimation for adaptive denoising and the threshold for early stopping in iterative learning.

\subsubsection{Adaptive Denoising Model} 
To effectively remove the noise in structural MRIs, we design a conditional UNet-like denoising network (see Fig.~\ref{figFramework}~(b)) that can adaptively predict the noise depending on our estimated noise levels. 
Specifically, this model contains five convolution blocks for MRI feature encoding and five convolution blocks for decoding,  
with each block containing two sequential convolution layers (kernel size: $3\times3\times3$) with batch normalization and LeakyReLU activation. 
The channel numbers of the five encoder blocks are 16, 32, 64, 128, and 128, respectively, while those for the five decoder blocks are 
128, 128, 64, 32, and 16, respectively. 
The feature maps in the encoder are downsampled by $2\times2\times2$ with max-pooling four times.
A previous study~\citep{odena2016deconvolution} has shown that upsampling using deconvolution may generate artifacts. 
So in this work, following~\citep{duffy2021retrospective}, we upsample the feature maps through linear interpolation in decoding.

A previous study~\citep{zhou2019unet++} has shown that redesigning skip connections of UNet~\citep{cciccek20163d} may promote exploiting multiscale features, alleviate network depth, and accelerate the inference speed.
Inspired by this work, we prune the original U-Net architecture to adapt it to our task at hand by removing some skip connections. 
Specifically, for the adaptive denoising model, we keep two skip connections of higher-resolution feature maps (generated by the first two convolution blocks), as shown in Fig.~\ref{figFramework} (b).

Before the final prediction layer, we normalize the feature maps with our estimated noise $\sigma_e$ as a condition using a conditional normalization layer~\citep{hou2022truncated}. 
As reported in~\citep{song2019generative}, defining ${\xi}/{\sigma^2}$ (where the Gaussian noise ${\xi}\sim \mathcal{N}(0, \sigma^2 \I)$) as the target of a denoising model can promote the accuracy of noise prediction when the noise level is small. 
Following~\citep{hou2022truncated}, we also train the adaptive denoising model to predict the noise divided by its variance. 
Then, the model output is further multiplied by $\sigma_e^2$ (\ie, the estimated noise variance of the input image) as the final predicted noise.
The final noise prediction is subtracted from the input MRI to get the denoised image. 

For this adaptive denoising model, we use the $l_1$-norm as the loss function: 
\begin{equation}
\label{eq_ln}
\small
 \L_{n} = ||\hat{\n}-{\xi}||_1, 
\end{equation}
where $\hat{\n}=\f_D(\x)$ denotes the final predicted noise image, $\xi$ is the manually added Gaussian noise, and $\x$ is the input image. 

During the training of this denoising model, we manually add Gaussian noise to each clean MRI, where the std is randomly sampled within [0.01, 0.30] as in~\citep{hou2022truncated}.   
The perturbed image (\ie, MRI + Gaussian Noise) is treated as input, and the Gaussian noise image is regarded as the ground-truth output. 
During inference, we will first estimate the noise std based on gradient map of an input MRI, and then use this adaptive denoising model to predict the noise image based on estimated noise levels.

\subsubsection{Anti-artifact Model}
As shown in Fig.~\ref{figFramework}~(c), the anti-artifact model also uses a pruned UNet-like architecture, where a skip connection of the highest feature map is removed. 
This is expected to restore a clean MRI image by filtering artifacts at low resolution. 
The other network hyperparameters are set as the same as those used in the denoising model. 
Following~\citep{duffy2022retrospective}, we utilize an $l_1$-norm loss to encourage the similarity between the predicted image and its ground truth (\ie, motion-free MRI), formulated as: 
\begin{equation}
\label{eq_lm}
\small
\L_{m} = ||\m-\hat{\m}||_1,
 \end{equation}
where $\hat{\m}$=$\f_A({\x})$ denotes the motion-corrected MRI estimation, and $\m$ is the matched ground-truth motion-free MRI. 

To retain brain anatomy details during motion correction,  
we further introduce a gradient-based loss function in the proposed anti-artifact model.
This will encourage the model not to distort the original brain structures in 3D MR images.
The gradient-based loss is formulated as follows: 
\begin{equation}
\label{eq_lg}
\small
 %\L_{g} = \frac{1}{D}\sum\nolimits_{i=1}^{D}||\bigtriangledown{\m_i}-\bigtriangledown \hat{\m}_i||_1,
 \L_{g} = ||\bigtriangledown{\m}-\bigtriangledown \hat{\m}||_1,
\end{equation}
Then, the total loss for the proposed anti-artifact model training can be written as:
\begin{equation}
\label{eq_a}
\small
 \L_{A} = \L_{m} + \L_{g}.
\end{equation}

\section{Experiments}
\label{S4}
%In this section, We first evaluate the proposed JDAC and visualize representative results of T1w MRIs with head motion in the MR-ART dataset. Secondly, We apply the trained JDAC on noised structural MRIs that are affected by motion artifacts in a clinical dataset related to depression study.  

\subsection{Experimental Setup} 
\textbf{Data \& Pre-processing}. 
Three datasets with T1-weighted MRI scans are employed, including ADNI, MR-ART, and NBOLD. 
All MRIs from the three
datasets are minimally preprocessed, including (1) skull stripping, and (2) intensity normalization to the range of [0, 1]. 
More details on the data are introduced as follows. 

(1) \textbf{ADNI}. 
A total of 9,544 T1-weighted MRI scans are downloaded from ADNI. 
These data are used to pretrain and validate denoising models with manually added Gaussian noise (\ie, 7,635 MRI scans for training, and the remaining 1,909 scans for validation). 

(2) \textbf{MR-ART}. 
The brain MRIs in this dataset are collected from $148$ healthy subjects. 
For each subject, one clean (without noise) and motion-free MRI scan (denoted as $\z$) and two motion-affected MRI scans (denoted as $\tilde{\z}$) of different artifact levels are acquired. 
This results in a total of 296 paired samples, with each pair containing one motion-affected MRI treated as input and its matched motion-free MRI as ground truth.  
With these matched images, we can quantitatively evaluate the performance of competing methods and our JDAC. 
A 5-fold subjective-level cross-validation strategy is adopted to avoid any bias introduced by random splitting.1) During \emph{training}, the matched MRI scans (\ie, $\tilde{\z}$) are used to train the anti-artifact model of JDAC.
For the training/fine-tuning of competing methods, the Gaussian noise with randomly selected std in the range of [0.01, 0.30] is added to each $\tilde{\z}$ to generate training samples (denoted as $\hat{\z}_{tr}$).
2) During \emph{test}, 
the Gaussian noise with a fixed std value is added to $\tilde{\z}$ for generating test samples (denoted as $\hat{\z}_{te}$).  
For inference, we input $\hat{\z}_{te}$ to a model and compare the output image $\hat{\z}^\prime$ of this model with its ground-truth original clean and motion-free MRI ${\z}$. 

(3) \textbf{NBOLD}. 
The T1-weighted MRI scans in NBOLD study are collected from $309$ subjects, where $30$ images are affected by noise and head motion. 
Since there are no ground-truth clean and motion-free images, we use these $30$ MRI scans for qualitative assessment of the proposed method.

\textbf{Evaluation Metric}. 
We evaluate the \emph{denoising performance} of JDAC and each competing method using four standard metrics, including (1) peak-signal-to-noise ratio (PSNR), (2) root mean square error (RMSE), (3) structural similarity index measure (SSIM), and (4) Multi-scale SSIM (MS-SSIM). 
Note that there are few established metrics to evaluate anti-artifact performance for real MRIs with motion artifacts and noise. 
Through experiments, we find that a gradient map is not sensitive to intensity distribution of its corresponding MR image, and helps preserve image texture information including artifacts (as shown in Fig.~\ref{fig_NoiseLevels}~(b)).  
Accordingly, for evaluating \emph{anti-artifact performance}, we propose to calculate the above four metrics on gradient maps of each motion-corrected MRI and its ground-truth motion-free image. 

%\if false
\begin{table*}[!t]
\setlength{\belowdisplayskip}{0pt}
\setlength{\abovedisplayskip}{0pt}
\setlength{\abovecaptionskip}{0pt}
\setlength{\belowcaptionskip}{0pt}
\scriptsize
%\tiny
\centering
\renewcommand{\arraystretch}{0.8}
\caption{Performance of seven methods in joint denoising and motion artifacts correction task on MR-ART dataset.
}
\label{comparison_MRART}
\setlength{\tabcolsep}{7pt}
\begin{tabular}{l|cccc|cccc}
\toprule
\multirow{2}{*}{Method} &\multicolumn{4}{c|}{Metrics on Image for Denoising Task} &\multicolumn{4}{c}{Metrics on Gradient Map for Anti-Artifact Task}\\
\cmidrule(lr){2-5} \cmidrule(lr){6-9} 
&PSNR (dB) &RMSE  &SSIM &MS-SSIM  &PSNR (dB) &RMSE  &SSIM  &MS-SSIM\\
\midrule
{DRN-DCMB}    
&22.84±1.02&0.0726±0.0090&0.6487±0.0418&0.9588±0.0174 &26.00±0.49&0.0502±0.0028&0.5317±0.0433&0.9208±0.0181\\
{SUNet}    
&25.56±1.23&0.0533±0.0082&0.8842±0.0276&0.9684±0.0126 &29.92±0.69&0.0320±0.0026&0.7627±0.0338&0.9502±0.0143\\
{BM4D}    
&24.88±2.33&0.0599±0.0274&0.4728±0.0505&0.9676±0.0440 &30.66±1.24&0.0296±0.0051&0.7821±0.0559&0.9539±0.0294\\
{UNet3D}    
&26.01±1.62&0.0510±0.0108&\textbf{0.8896±0.0365}&0.9752±0.0159 &29.99±0.78&0.0318±0.0029&{0.7804±0.0382}&\textbf{0.9579±0.0157}\\
{nnUNet}    
&25.96±1.44&0.0511±0.0094&0.8830±0.0304&\textbf{0.9774±0.0133} &28.24±0.61&0.0388±0.0028&0.7316±0.0364&0.9527±0.0159\\
{FONDUE}    
&24.29±1.04&0.0615±0.0080&0.8529±0.0300&0.9657±0.0130 &27.26±0.52&0.0435±0.0026&0.6861±0.0332&0.9456±0.0154\\
\midrule
{JDAC~(Ours)}
&\textbf{26.46±1.26}&\textbf{0.0480±0.0071}&{0.8690±0.0287}&0.9591±0.0160 &\textbf{33.07±1.10}&\textbf{0.0224±0.0028}&\textbf{0.7930±0.0273}&0.9550±0.0095\\
\bottomrule
\end{tabular}
\end{table*}
%\fi 

\textbf{Implementation}. 
The adaptive denoising model and the anti-artifact model in our JDAC are trained in a sequential manner. 
(1) We first train the denoiser on MRIs from ADNI, with manually added Gaussian noise as supervision, and train the anti-artifact model using MRIs in the MR-ART dataset, with matched motion-free MRIs as supervision. 
An Adam optimizer~\citep{kingma2014adam} is used in these two models, with a learning rate of ${10}^{-4}$, and batch size of 2. 
The training epoch for the adaptive denoising model is set as 15 with {\color{black}7K+} training MRIs. 
We run the anti-artifact model training (with {\color{black}~441} MRIs of 147 subjects) for 150 epochs due to the relatively small data size. 
Both models are trained with randomly selected image patches (size $128 \times 128 \times 128$) as inputs using PyTorch with NVIDIA TITAN Xp (memory: 12GB).
(2) We then use the proposed iterative framework with the trained models to jointly perform denoising and motion artifact correction on test data with an early stopping strategy applied.  
In the test, a whole MRI volume is fed into a model on a CPU platform (\ie, Intel(R) Core (TM) i7-8700K CPU @ 3.70GHz) with 64GB RAM due to the limited GPU memory.

\subsection{Competing Methods}
We compare JDAC with two 2D-based methods (\ie, DRN-DCMB and SUNet) and four 3D-based state-of-the-art methods (\ie, BM4D, UNet3D, nnUNet, and FONDUE) for denoising or motion correction.
Among them, the BM4D is a traditional method and the others are deep learning models.
The details of the competing methods are introduced below.

(1) \textbf{DRN-DCMB}~\citep{liu2020motion}: 
The DRN-DCMB is a residual CNN with densely connected multi-resolution blocks to predict a residual image and reduce motion artifacts in T1-weighted MRIs acquired at different imaging planes.
Since DRN-DCMB is a 2D-based model, we train this model with randomly selected slices of MRIs in MR-ART with motion artifacts and simulated Gaussian noise.  
The trained model is applied to MRI volumes slice-by-slice during inference. % testing.

(2) \textbf{SUNet}~\citep{al2022stacked}: 
The SUNet is an eﬃcient retrospective 2D method using stacked UNets to address the problem of rigid motion artifacts.
Specifically, it first employs a UNet to learn structural details from adjacent slices for prediction and then uses another UNet to preserve spatial structure details and refine the pixel-to-pixel prediction.
Similar to the 2D DRN-DCMB method, we train and test the model in the same way as DRN-DCMB.
 
(3) \textbf{BM4D}:
BM4D is a 4D implementation of the popular denoising model BM3D~\citep{dabov2007image} based on~\citep{makinen2020collaborative, makinen2022ring} to reduce additive spatially correlated stationary Gaussian noise for 3D volumetric data.
We use a Python package of BM4D binaries\footnote{https://pypi.org/project/bm4d/} for implementation and directly apply BM4D to test MRIs for comparison. 

(4) \textbf{UNet3D}~\citep{cciccek20163d}: The UNet is one of the most popular architectures in medical image denoising~\citep{lehtinen2018noise2noise}, anti-artifact~\citep{zhang2018ct}, and restoration~\citep{liu2022udc,hephzibah2023review}.
In the experiments, we use the 3D implementation of UNet from MONAI\footnote{https://docs.monai.io/en/stable/networks.html\#basicunet} as the baseline of 3D deep learning models. 

(5) \textbf{nnUNet}~\citep{isensee2019no}: The nnUNet is similar to UNet3D, but has a modified network architecture. 
Following~\citep{duffy2021retrospective}, the nnUNet uses trilinear upsampling instead of 
deconvolution upsampling and batch normalization rather than instance normalization for artifact reduction.
The nnUNet is trained following the strategy used in the previous study~\citep{duffy2021retrospective}.

(6) \textbf{FONDUE}~\citep{adame2023fondue}: 
The FONDUE is a deep CNN model designed for denoising multi-resolution structural MRI.
This method is trained with diverse MRIs from different datasets and has resolution-invariant capabilities. 
Considering the parameter scale of FONDUE and the limitation in our GPU memory, we download the pre-trained model following instructions of the open source code on GitHub\footnote{https://github.com/waadgo/FONDUE} and fine-tune it on the MR-ART dataset.

%%%%%%%%%%%%%%%%%%%%%%%%%%%%%%%%%%%%%%%%%%%%%%%
\begin{figure*}[!t]
\setlength{\belowdisplayskip}{0pt}
\setlength{\abovedisplayskip}{0pt}
\setlength{\abovecaptionskip}{-2pt}
\setlength{\belowcaptionskip}{0pt}
\centering
\includegraphics[width=1.0\textwidth]{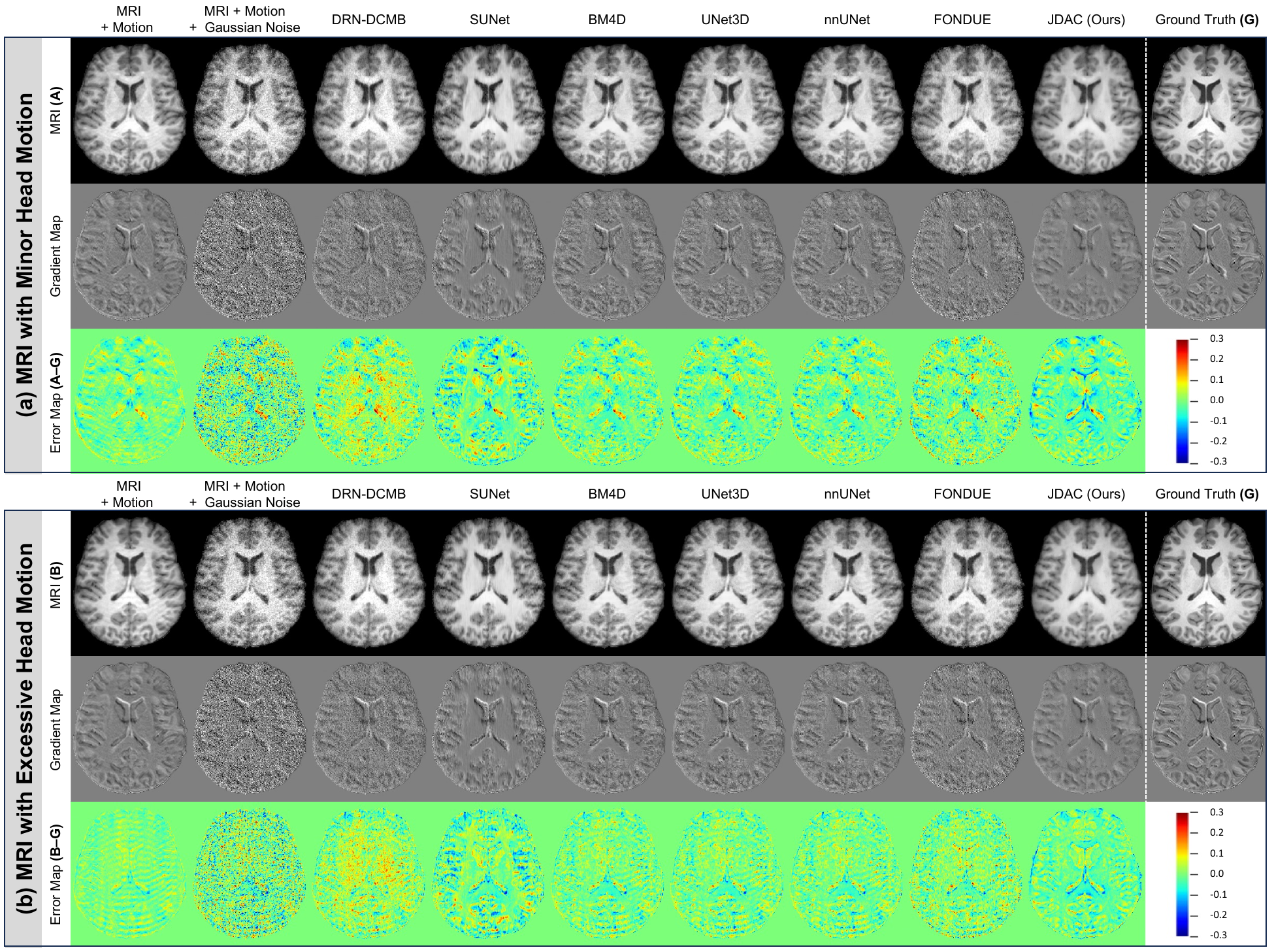}
\caption{Qualitative comparison of the JDAC against the six competing methods (\ie, DRN-DCMB, SUNet, BM4D, UNet3D, nnUNet, and FONDUE) on one subject (ID: 862915) in the MR-ART~\citep{narai2022movement} dataset with manually added Gaussian noise (std: 0.10). The panel (a) shows the denoising and anti-artifact results of MRI with minor head motion, and the panel (b) shows results of the same subject with excessive head motion. 
}
\label{fig_vasualize_MRART}
\end{figure*}

Note that, the BM4D, UNet3D, and nnUNet methods are initially designed for medical image denoising and others are developed for motion artifact correction.
To ensure a fair comparison, we make our strategies for training learning-based models, other than BM4D, as consistent as possible with the strategies used for training the denoising and anti-artifact models in our JDAC framework. 
That is, we use the training set of ADNI to pre-train each competing deep learning model with the same noise generation strategy as the training of the denoising model in our JDAC. 
The training data from MR-ART is then used for fine-tuning the above-pretrained model for artifact correction, which is also similar to the training of the anti-artifact model in JDAC. 
Since the competing deep models need to perform both image denoising and motion artifact correction, we also added Gaussian noise to the training MRIs (\ie, $\hat{\z}_{tr}$) from the MR-ART dataset for fine-tuning. 
The details are introduced in the description of the MR-ART dataset. 
In the experiments, we typically use the default setting of all competing methods and make a concerted effort to ensure that the network architecture and hyperparameters are comparable to the proposed JDAC.

%%%%%%%%%%%%%%%%%%%%%%%%%%%%%%%%%%%%%%%%%%%%%%%%%%%%%%%%%%%%%%%%%%
\subsection{Joint Denoising and Motion Correction on MR-ART}
We report the results achieved by the proposed JDAC and the competing methods on test MRIs from MR-ART for joint denoising and motion artifact correction % of the MR-ART data 
in Table~\ref{comparison_MRART}.  
From Table~\ref{comparison_MRART}, we have the following observations. 

\emph{First}, our JDAC generally outperforms the competing methods in most cases. 
For instance, JDAC achieves the best PSNR ($33.07$dB), RMSE ($0.0224$), and SSIM values ($0.7930$) in the task of motion correction, while the second-best results are achieved by UNet3D (with  PSNR: $29.99$ dB, RMSE: $0.0318$), and SSIM: $0.7804$).
Meanwhile, the left part of Table~\ref{comparison_MRART} suggests that JDAC achieves the best PSNR and RMSE values (\ie, $26.46$dB, and $0.0480$, respectively) for denoising. 
It should be noted that the six competing methods do not account for the relationship between denoising and motion correction tasks. 
Contrastly, JDAC utilizes a joint learning strategy for the two tasks, which might be a key factor in explaining the superior performance of our method. 
\emph{Additionally}, 3D methods (\ie, BM4D, UNet3D, and nnUNet, and our JDAC) generally outperform 2D methods (\ie, DRN-DCMB, and SUNet) in the two tasks.  
This may be due to the fact that 3D methods can make use of spatial volumetric information when compared with 2D methods. 
\emph{Furthermore}, among the six competing methods, UNet3D usually produces the best results in both tasks of denoising and motion correction. 
This indicates that when both tasks need to be processed simultaneously, a universal CNN model may perform better than specially designed denoising or artifact correction neuro networks, which means ignoring the impact between these two tasks will reduce the processed image quality.

%%%%%%%%%%%%%%%%%%%%%%%%%%%%%%%%%%%%%%%%%%%%%%%
\begin{figure*}[!t]
\setlength{\belowdisplayskip}{0pt}
\setlength{\abovedisplayskip}{0pt}
\setlength{\abovecaptionskip}{0pt}
\setlength{\belowcaptionskip}{0pt}
\centering
\includegraphics[width=1.0\textwidth]{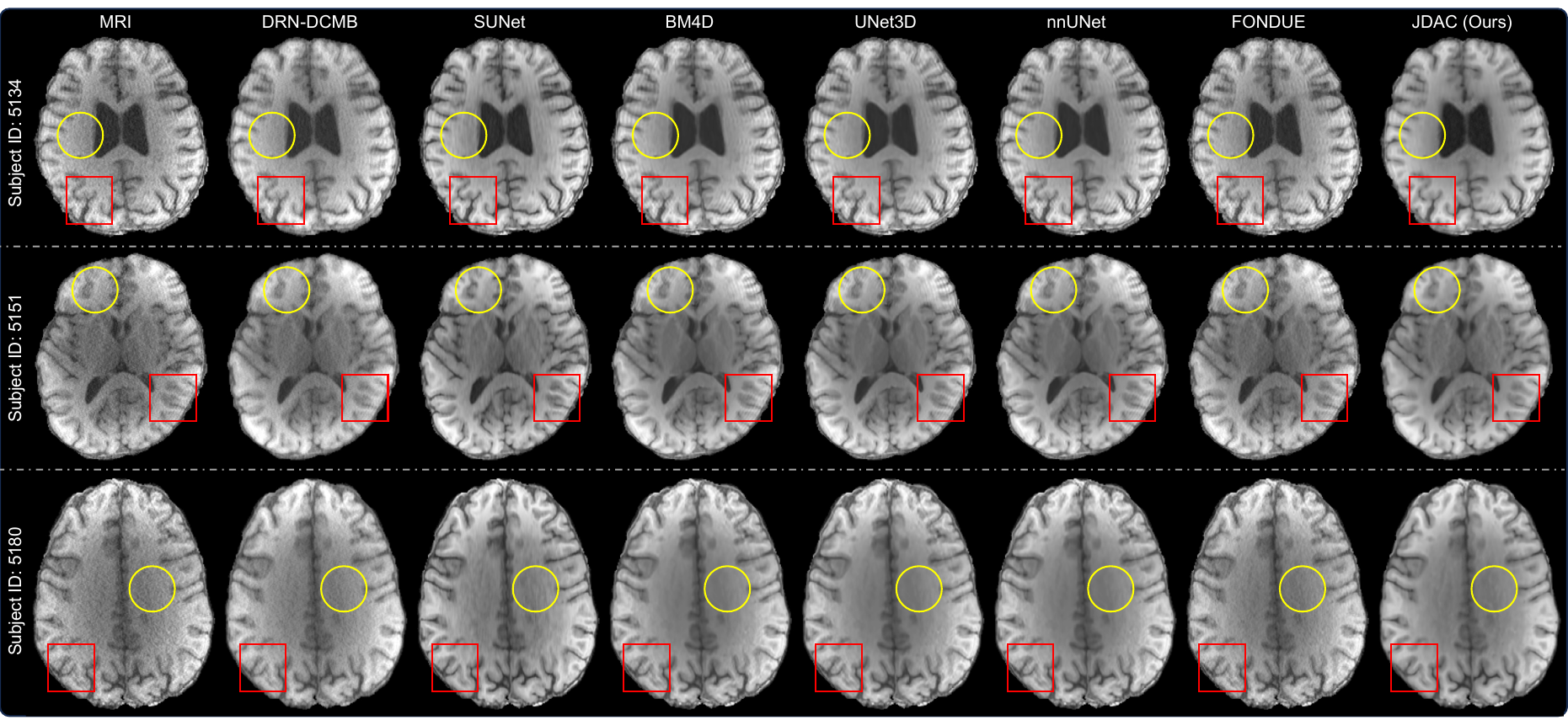}
\caption{Real motion cases of three subjects in the NBOLD study~\citep{steffens2017negative}. 
The first column images are the motion-affected real clinical MRIs, while the other columns are the denoised and motion-corrected images via the competing methods and the proposed JDAC. The circles show areas with significantly different denoising results, and the rectangles highlight areas with different artifact correction results. }
\label{fig_vasualize_NBOLD}
\end{figure*}
%%%%%%%%%%%%%%%%%%%%%%%%%%%%%%%%%%%%%%%%%%%%%%%

Apart from the quantitative results in Table~\ref{comparison_MRART}, we further visualize the results of our JDAC and each competing method for MRIs of one typical test subject with different motion severity from MR-ART in Fig.~\ref{fig_vasualize_MRART}. 
From Fig.~\ref{fig_vasualize_MRART}, we have several interesting observations. 
\emph{First}, our JDAC outperforms the competing methods by visually comparing the corrected MRI slices in the first row of both Fig.~\ref{fig_vasualize_MRART} (a) and Fig.~\ref{fig_vasualize_MRART} (b). 
By comparing the corrected MRI results of JDAC in the two sub-figures, we can observe that JDAC can obtain comparable image quality improvements on MR images with minor and excessive motion artifacts. 
This may imply that our JDAC can be well applied to MRI images with different motion severity.  
\emph{Second}, comparing the gradient maps in the second rows of both subpanels in Fig.~\ref{fig_vasualize_MRART}, we can see our JDAC can output a cleaner gradient map than competing methods, especially in the brain cerebral cortex regions.
This implies that JDAC is able to keep more structural information than others in the denoising task. 
\emph{Besides}, from error maps in Fig.~\ref{fig_vasualize_MRART}, one can easily observe the severity of motion artifacts and check the motion correction results achieved by a specific method. 
The error maps in Fig.~\ref{fig_vasualize_MRART} (a) show that all methods can handle MRIs with minor motion artifacts well.
\emph{In addition}, error maps of Fig.~\ref{fig_vasualize_MRART} (b) demonstrate that only JDAC and SUNet can remove excessive motion artifacts more effectively, compared with the other methods. 
In particular, it can be seen that JDAC can output MRI with clear brain anatomy structures, but SUNet generates a more blurred MRI image, especially in the cerebral cortex regions with complex structures.  
These results imply that the proposed iterative learning strategy in JDAC can remove excessive motion artifacts effectively. 
\emph{Furthermore}, from the second rows of Fig.~\ref{fig_vasualize_MRART}~(a) and (b), we can see that the gradient map of MR images is typically sensitive to noise. 
This observation further supports the rationality of our proposed method for estimating image noise levels using gradient maps. 

%%%%%%%%%%%%%%%%%%%%%%%%%%%%%%%%%%%%%%%%%%%%%%%
\begin{figure}[!t]
\setlength{\belowdisplayskip}{0pt}
\setlength{\abovedisplayskip}{0pt}
\setlength{\abovecaptionskip}{0pt}
\setlength{\belowcaptionskip}{0pt}
\centering
\includegraphics[width=0.48\textwidth]{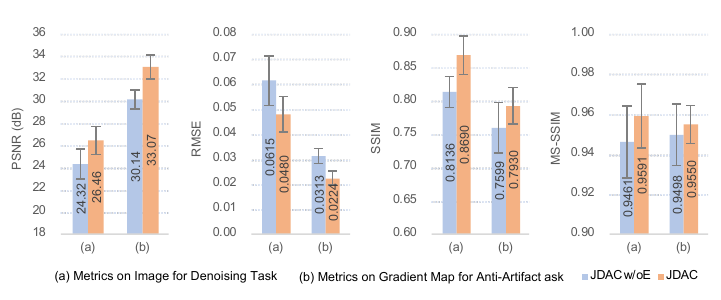}
\caption{Performance of JDAC and its variants with \& without noise estimation strategy %({\ie, JDAC and JDACw/oE}) 
for denoising and motion artifact correction 
on MR-ART.}
\label{fig_ablation_without_estimation}
\end{figure}
%%%%%%%%%%%%%%%%%%%%%%%%%%%%%%%%%%%%%%%%%%%%%%%

%%%%%%%%%%%%%% New Section %%%%%%%%%
\subsection{Application to NBOLD Study} 
To evaluate the generalization ability of our JDAC to different studies, we further apply it to the independent NBOLD study~\citep{steffens2017negative}. 
In this study, the 30 MRI scans are corrupted by both motion artifacts and noise, but there is no ground truth (\ie, motion-free and clear images). 
So we directly apply our JDAC and the competing methods trained on ADNI and MR-ART to MRIs in NBOLD and provide several cases in Fig.~\ref{fig_vasualize_NBOLD} for qualitative visual comparison. 
From Fig.~\ref{fig_vasualize_NBOLD}, we have similar observations to those in Fig.~\ref{fig_vasualize_MRART}, that is, the proposed JDAC produces overall better results in most cases. 

In Fig.~\ref{fig_vasualize_NBOLD}, we highlight several internal homogeneous regions of white matter (WM) or gray matter (GM) in the brain using yellow circles. 
For the WM regions, the BM4D, UNet3D, nnUNet, and JDAC remove the image noise better than others (see circles on images of subjects 5134 and 5180). 
Meanwhile, JDAC can well keep the boundary of GM and WM regions in MRI of subject 5151, but other methods often produce blurry results. 
This further verifies that the proposed method helps preserve brain anatomical structure during denoising and motion artifact correction. 
{Additionally}, we use red boxes to emphasize a local region that shows notable differences among each of the seven methods for motion correction in Fig.~\ref{fig_vasualize_NBOLD}. 
From the corrected MRI slices in Fig.~\ref{fig_vasualize_MRART}, we can find that JDAC achieves good motion correction performance in MRI regions with complex structures (marked by red boxes), while the other methods often retain a large amount of motion artifacts. 
These results further validate that jointly performing denoising and motion artifact correction in an iterative learning framework can promote denoising and artifact correction performance.

%%%%%%%%%%%%%%%%%%%%%%%%%%%%%%%%%%%%%%%%%%%%%%%
\begin{table}[!t]
\setlength{\belowdisplayskip}{0pt}
\setlength{\abovedisplayskip}{0pt}
\setlength{\abovecaptionskip}{0pt}
\setlength{\belowcaptionskip}{0pt}
\scriptsize
%\tiny
\centering
\renewcommand{\arraystretch}{0.9}
\caption{Performance of JDAC and its two variants in joint denoising and motion artifacts correction task on MR-ART dataset.}
\label{tab_ablation}
\setlength{\tabcolsep}{5pt}
\begin{tabular}{l|cccc}
\toprule
\multirow{2}{*}{Method} &\multicolumn{4}{c}{Metrics on Image for Denoising Task}\\
\cmidrule(lr){2-5}
&PSNR (dB) &RMSE  &SSIM &MS-SSIM \\
\midrule
{JDACw/oA} &\textbf{26.50±1.52}&{0.0481±0.0093}&\textbf{0.8952±0.0296}&\textbf{0.9774±0.0130} \\
{JDACw/oD} &20.90±0.64&0.0903±0.0065&0.7680±0.0241&0.9166±0.0132\\
{JDAC~(Ours)}&{26.46±1.26}&\textbf{0.0480±0.0071}&{0.8690±0.0287}&0.9591±0.0160 \\
\toprule
{Method} &\multicolumn{4}{c}{Metrics on Gradient Map for Anti-Artifact Task}\\
\midrule
{JDACw/oA}    &29.53±0.61&0.0335±0.0024&0.7481±0.0346&\textbf{0.9554±0.0153}\\
{JDACw/oD}    &29.34±0.57&0.0342±0.0022&0.6826±0.0289&0.9116±0.0114\\
{JDAC~(Ours)}&\textbf{33.07±1.10}&\textbf{0.0224±0.0028}&\textbf{0.7930±0.0273}&0.9550±0.0095\\
\bottomrule
\end{tabular}
\end{table}
%%%%%%%%%%%%%%%%%%%%%%%%%%%%%%%%%%%%%%%%%%%%%%%

%%%%%%%%%%%%%%%%%%%%%%%%%%%%%%%%%%%%%%%%%%%%%%%%%%%%%%%%%%%%%%%%%%%%%%%%%%%%%%%%%%%%%%%%%%%%%%%%%%%%%%%%%%%%%%%%%%%%%%%%%%
%\if false
\begin{table*}[!t]
\setlength{\belowdisplayskip}{0pt}
\setlength{\abovedisplayskip}{0pt}
\setlength{\abovecaptionskip}{0pt}
\setlength{\belowcaptionskip}{0pt}
\scriptsize
%\tiny
\centering
\renewcommand{\arraystretch}{0.8}
\caption{Results of JDAC with different %influence of 
iteration steps in joint denoising and motion artifact correction task on MR-ART dataset. 
}
\label{ablation_iterations}
\setlength{\tabcolsep}{6.5pt}
\begin{tabular*}{1\textwidth}{l|c|cccc|cccc}
\toprule
\multirow{2}{*}{Noise Std} &\multirow{2}{*}{Iterations} &\multicolumn{4}{c|}{Metrics on Image for Denoising Task} &\multicolumn{4}{c}{Metrics on Gradient Map for Anti-Artifact Task}\\
\cmidrule(lr){3-6} \cmidrule(lr){7-10} 
&&PSNR (dB) &RMSE  &SSIM &MS-SSIM  &PSNR (dB) &RMSE  &SSIM  &MS-SSIM\\
\midrule
\multirow{4}{*}{0.025}
 &{\color{black}1}&\textbf{24.92±2.45}&\textbf{0.0592±0.0184}&0.8769±0.0283&0.9481±0.0252 &32.93±1.49&0.0229±0.0038&0.8054±0.0348&0.9565±0.0104\\
 &2&24.43±3.18&0.0644±0.0257&\textbf{0.8913±0.0291}&\textbf{0.9508±0.0302} &33.66±1.52&0.0211±0.0036&0.8241±0.0332&0.9621±0.0096\\
 &3&22.92±3.72&0.0785±0.0357&0.8877±0.0340&0.9478±0.0339 &34.23±1.59&0.0198±0.0035&0.8356±0.0335&0.9653±0.0095\\
 &4&22.89±3.42&0.0775±0.0309&0.8900±0.0318&0.9490±0.0319 &\textbf{34.33±1.55}&\textbf{0.0195±0.0034}&\textbf{0.8384±0.0331}&\textbf{0.9660±0.0095}\\
\midrule
\multirow{4}{*}{0.050}
 &1&25.73±1.80&0.0529±0.0118&0.8680±0.0280&0.9525±0.0196 &32.69±1.42&0.0235±0.0038&0.7920±0.0335&0.9545±0.0106\\
 &2&\textbf{25.94±2.21}&\textbf{0.0523±0.0158}&0.8777±0.0276&\textbf{0.9540±0.0222} &33.17±1.54&0.0223±0.0038&0.8053±0.0339&0.9576±0.0101\\
 &3&24.09±3.81&0.0695±0.0356&0.8822±0.0345&0.9506±0.0313 &33.85±1.65&0.0206±0.0037&0.8245±0.0356&0.9630±0.0102\\
 &4&24.40±3.58&0.0660±0.0303&\textbf{0.8849±0.0320}&0.9529±0.0286 &\textbf{33.92±1.66}&\textbf{0.0205±0.0038}&\textbf{0.8251±0.0349}&\textbf{0.9635±0.0102}\\
\midrule
\multirow{4}{*}{0.075}
 &1&26.14±1.27&0.0498±0.0073&0.8545±0.0266&0.9545±0.0150 &32.30±1.36&0.0246±0.0038&0.7724±0.0323&0.9508±0.0108\\
 &2&\textbf{26.44±1.32}&\textbf{0.0482±0.0074}&0.8636±0.0273&0.9550±0.0173 &32.79±1.36&0.0232±0.0035&0.7878±0.0310&0.9537±0.0101\\
 &3&25.00±2.28&0.0583±0.0163&\textbf{0.8836±0.0283}&0.9573±0.0210 &\textbf{33.73±1.48}&\textbf{0.0209±0.0034}&\textbf{0.8163±0.0324}&0.9614±0.0101\\
 &4&25.36±2.21&0.0558±0.0150&0.8817±0.0282&\textbf{0.9594±0.0198} &33.69±1.41&0.0210±0.0033&0.8138±0.0307&\textbf{0.9616±0.0100}\\
\midrule
\multirow{4}{*}{0.100}
 &1&25.99±1.21&0.0507±0.0073&0.8358±0.0256&0.9524±0.0132 &31.70±1.24&0.0263±0.0037&0.7470±0.0302&0.9445±0.0109\\
 &2&26.34±1.00&0.0485±0.0059&0.8403±0.0276&0.9519±0.0150 &32.15±1.21&0.0249±0.0034&0.7612±0.0289&0.9466±0.0101\\
 &3&\textbf{26.46±1.26}&\textbf{0.0480±0.0071}&\textbf{0.8690±0.0287}&0.9591±0.0160 &\textbf{33.07±1.10}&\textbf{0.0224±0.0028}&\textbf{0.7930±0.0273}&0.9550±0.0095\\
 &4&26.20±1.47&0.0497±0.0086&0.8666±0.0284&\textbf{0.9592±0.0168} &33.06±1.15&\textbf{0.0224±0.0029}&0.7909±0.0282&\textbf{0.9556±0.0100}\\
\midrule
\multirow{4}{*}{0.125}
 &1&25.40±1.23&0.0542±0.0078&0.8148±0.0252&0.9478±0.0131 &30.96±1.10&0.0285±0.0035&0.7205±0.0289&0.9364±0.0109\\
 &2&25.85±1.02&0.0514±0.0063&0.8213±0.0277&0.9478±0.0150 &31.48±1.06&0.0269±0.0032&0.7354±0.0279&0.9389±0.0103\\
 &3&\textbf{26.26±1.42}&\textbf{0.0493±0.0084}&\textbf{0.8521±0.0294}&\textbf{0.9569±0.0160} &32.62±1.04&0.0236±0.0028&0.7741±0.0280&0.9500±0.0101\\
 &4&25.89±1.65&0.0517±0.0104&0.8507±0.0295&0.9568±0.0165 &\textbf{32.76±1.01}&\textbf{0.0232±0.0026}&\textbf{0.7763±0.0263}&\textbf{0.9517±0.0095}\\
\midrule
\multirow{4}{*}{0.150}
 &1&24.54±1.16&0.0598±0.0081&0.7901±0.0249&0.9410±0.0132 &30.02±1.00&0.0318±0.0036&0.6917±0.0286&0.9258±0.0112\\
 &2&25.06±0.88&0.0561±0.0059&0.7950±0.0268&0.9408±0.0148 &30.54±0.91&0.0299±0.0031&0.7036±0.0264&0.9278±0.0102\\
 &3&25.65±1.08&0.0526±0.0067&\textbf{0.8264±0.0285}&0.9519±0.0151 &\textbf{31.83±0.84}&\textbf{0.0257±0.0025}&\textbf{0.7450±0.0253}&0.9416±0.0098\\
 &4&\textbf{25.92±1.18}&\textbf{0.0511±0.0071}&0.8255±0.0276&\textbf{0.9526±0.0149} &31.81±0.99&0.0258±0.0029&0.7433±0.0261&\textbf{0.9419±0.0100}\\
\bottomrule
\end{tabular*}
\end{table*}
%\fi
%%%%%%%%%%%%%%%%%%%%%%%%%%%%%%%%%%%%%%%%%%%%%%%%%%%%%%%%%%%%%%%%%%%%%%%%%%%%%%%%%%%%%%%%%%%%%%%%%%%%%%%%%%%%%%%%%%%%%%%%%%

%%%%%%%%%%%%%%%%%%%%%%%%%%%% new section   new section  new section %%%%%%%%%%%%%%%%%%%%%%%%%%%%%%%%%%%
\section{Discussion}
\label{S5}
\if false
In this section, we analyze and discuss the influences of several important components of the proposed JDAC method. 
We also discuss the limitations of the current work and present several future research directions. 
\fi 

%%%%%% - New Subsection - %%%%%%

\subsection{Influence of Noise Level Estimation}
We compare our JDAC with its degraded variant that does not use noise level estimation strategy, that is the denoising model in Fig.~\ref{figFramework} (b) without conditional normalization layer (called \textbf{JDACw/oE}).
The variant model is also pretrained as the adaptive denoising model and applied in the iterative learning framework of JDAC.
The results shown in Fig.~\ref{fig_ablation_without_estimation} demonstrate that JDAC outperforms JDACw/oE on all evaluation metrics with both denoising and anti-artifact tasks.
This implies that the noise level estimation strategy is essential in the proposed iterative learning framework. 
This also indicates that the improvement in noise removal can significantly promote the effect of artifact correction.
Thus in the following ablation studies, we keep the noise level estimation strategy for the other variants of JDAC.

%%%%%%%%%%%%%%%%%%%%%%%%%%%%%%%%%%%%%%%%%%%%%%

%%%%%% - New Subsection - %%%%%%
%%%%%%%%%%%%%%%%%%%%%%%%%%%%%%%%%%%%%%%%%%%%%%%
%\subsection{Ablation Study}
\subsection{Influence of Joint Denoising and Artifact Correction}
To validate the effectiveness of joining two tasks of denoising and anti-artifact. 
we compare the JDAC with its other two variants that only perform denoising (\ie, the denoiser called \textbf{JDACw/oA}) or anti-artifact (\ie, the anti-artifact model called \textbf{JDACw/oD}) alone. 
For instance, JDACw/oA will feed the denoised image output by the adaptive denoising model at the $k$-th iteration back into itself at the next iteration.
The same iterative learning strategy is adapted for JDACw/oD.
The same early stopping strategy of JDAC is used in these two variants. 

As shown in Table~\ref{tab_ablation}, the JDAC performs better than its variants in metrics of the gradient maps. 
Meanwhile, JDAC can achieve almost the same performance as the JDACw/oA after the iteration in metrics of the images themselves. 
This implies that the iterative algorithm helps improve the motion artifact correction performance under the impact of Gaussian noise. 
Also, the JDACw/oA performs much better than JDACw/oD, implying that the performance of the anti-artifact model (\ie, JDACw/oD) is more susceptible to Gaussian noise.

%%%%%% - New Subsection - %%%%%%
%%%%%%%%%%%%%%%%%%%%%%%%%%%%%%%%%%%%%%%%%%%%%%%%%%%%%%%%%%%%%%%%%%%%%%%%%%%%%%%%%%%%%%%%%%%%%%%%%%%%%%%%%%%%%%%%%%%%%%%%%%
\subsection{Influence of Iterative Learning}
\label{S_DiscussionIterative}
In this work, we use an early stopping strategy to accelerate the proposed iterative framework for joint denoising and anti-artifact.  
To investigate the influence of this early stopping strategy, we report the results of JDAC using different iteration steps (\ie, without using the early stopping strategy) on test data from MR-ART in Table~\ref{ablation_iterations}. 
The left part of Table~\ref{ablation_iterations} demonstrates that using more iterations does not invariably lead to improved denoising results.  
In most cases (\ie, with noise std $<0.150$), our method achieves the best performance using less than three iteration steps. 
The most possible reason is that the images tend to be more blurred rather than denoised with more iterations, so it is a better choice to decide the iterations depending on the image noise level. 
\emph{On the other hand}, as shown in the right part of Table~\ref{ablation_iterations}, the anti-artifact results tend to be better with more iterations under the influence of different noise levels.  
To balance the trade-off between denoising and anti-artifact, in this work, we develop an early stopping strategy based on the estimated image noise levels. 
When the estimated noise level falls below a certain threshold, it indicates over-denoising, prompting the need to halt the iterations to prevent further removal of subtle anatomical structures and motion artifacts in the brain.

%%%%%% - New Subsection - %%%%%%
%%%%%%%%%%%%%%%%%%%%%%%%%%%%%%%%%%%%%%%%%%%%%%%%%%%%%%%%%%%%%%%%%%%%%%%%%%%%%%%%%%%%%%%%%%%%%%%%%%%%%%%%%%%%%%%%%%%%%%%%%%%%%
\begin{table*}[!t]
\setlength{\belowdisplayskip}{0pt}
\setlength{\abovedisplayskip}{0pt}
\setlength{\abovecaptionskip}{0pt}
\setlength{\belowcaptionskip}{0pt}
%\footnotesize
\scriptsize
%\tiny
\centering
\renewcommand{\arraystretch}{0.8}
\caption{Results of JDAC with different noise levels in joint denoising and motion artifact correction task on MR-ART dataset. }
\label{ablation_noise_level}
\setlength{\tabcolsep}{6pt}
\begin{tabular*}{1\textwidth}{l|c|cccc|cccc}
\toprule
\multirow{2}{*}{Method} &\multirow{2}{*}{Noise Std} &\multicolumn{4}{c|}{Metrics on Image for Denoising Task} &\multicolumn{4}{c}{Metrics on Gradient Map for Anti-Artifact Task}\\
\cmidrule(lr){3-6} \cmidrule(lr){7-10} 
&&PSNR (dB) &RMSE  &SSIM &MS-SSIM  &PSNR (dB) &RMSE  &SSIM  &MS-SSIM\\
\midrule
{JDACw/oA}    
&  \multirow{3}{*}{$0.000$} &28.24±2.36&0.0402±0.0120&0.9311±0.0368&0.9828±0.0139&32.51±1.54&0.0241±0.0045&0.8492±0.0542&0.9699±0.0179\\
{JDACw/oD}    
& &28.80±2.20&0.0376±0.0106&0.9279±0.0326&\textbf{0.9851±0.0126}&\textbf{33.12±1.42}&\textbf{0.0224±0.0038}&\textbf{0.8678±0.0481}&\textbf{0.9743±0.0158}\\
{JDAC~(Ours)}    
& &\textbf{28.81±2.20}&\textbf{0.0375±0.0106}&\textbf{0.9408±0.0325}&\textbf{0.9851±0.0126}&33.00±1.35&0.0227±0.0037&0.8647±0.0468&0.9737±0.0156\\
\midrule

{JDACw/oA}    
&\multirow{3}{*}{$0.025$}  &27.73±2.08&0.0423±0.0113&0.9204±0.0345&0.9818±0.0137&31.11±1.04&0.0280±0.0036&0.8054±0.0456&0.9662±0.0172\\
{JDACw/oD}    
&&28.19±1.96&0.0400±0.0101&0.9178±0.0339&0.9830±0.0135&\textbf{32.56±1.22}&\textbf{0.0238±0.0035}&0.8482±0.0465&0.9703±0.0164\\
{JDAC~(Ours)}    
& &\textbf{28.31±1.98}&\textbf{0.0395±0.0100}&\textbf{0.9324±0.0325}&\textbf{0.9835±0.0129}&\textbf{32.56±1.21}&\textbf{0.0238±0.0035}&\textbf{0.8494±0.0457}&\textbf{0.9709±0.0158}\\
\midrule

{JDACw/oA}    
&\multirow{3}{*}{$0.050$}  &27.10±1.83&0.0452±0.0106&0.9080±0.0332&\textbf{0.9803±0.0136} &29.95±0.85&0.0320±0.0032&0.7692±0.0417&0.9613±0.0168\\
{JDACw/oD}    
& &25.58±1.24&0.0531±0.0080&0.8625±0.0317&0.9648±0.0144 &30.98±0.80&0.0284±0.0026&0.7733±0.0370&0.9466±0.0152\\
{JDAC~(Ours)}    
& &\textbf{27.43±1.74}&\textbf{0.0434±0.0097}&\textbf{0.9163±0.0329}&0.9801±0.0138 &\textbf{31.80±1.02}&\textbf{0.0259±0.0032}&\textbf{0.8219±0.0433}&\textbf{0.9649±0.0162}\\
\midrule

{JDACw/oA}    
&\multirow{3}{*}{$0.075$} &26.80±1.68&0.0466±0.0100&\textbf{0.9032±0.0318}&\textbf{0.9790±0.0134}&29.94±0.77&0.0320±0.0030&0.7688±0.0392&0.9588±0.0162\\
{JDACw/oD}    
& &22.65±0.92&0.0741±0.0078&0.8020±0.0255&0.9365±0.0132&29.87±0.61&0.0322±0.0022&0.7128±0.0296&0.9225±0.0117\\
{JDAC~(Ours)}    
& &\textbf{26.95±1.88}&\textbf{0.0461±0.0117}&0.9000±0.0345&0.9773±0.0152&\textbf{31.27±1.12}&\textbf{0.0276±0.0039}&\textbf{0.8042±0.0421}&\textbf{0.9609±0.0166}\\
\midrule

{JDACw/oA}    
&\multirow{3}{*}{$0.100$}  
&\textbf{26.50±1.52}&{0.0481±0.0093}&\textbf{0.8952±0.0296}&\textbf{0.9774±0.0130} &29.53±0.61&0.0335±0.0024&0.7481±0.0346&\textbf{0.9554±0.0153}\\
{JDACw/oD}    
& &20.90±0.64&0.0903±0.0065&0.7680±0.0241&0.9166±0.0132 &29.34±0.57&0.0342±0.0022&0.6826±0.0289&0.9116±0.0114\\
{JDAC~(Ours)}    
& 
&{26.46±1.26}&\textbf{0.0480±0.0071}&{0.8690±0.0287}&0.9591±0.0160 &\textbf{33.07±1.10}&\textbf{0.0224±0.0028}&\textbf{0.7930±0.0273}&0.9550±0.0095\\
\midrule

{JDACw/oA}    
&\multirow{3}{*}{$0.125$}  &\textbf{26.33±1.44}&\textbf{0.0489±0.0090}&\textbf{0.8931±0.0289}&\textbf{0.9757±0.0129}&29.87±0.65&0.0322±0.0025&0.7559±0.0343&\textbf{0.9536±0.0148}\\
{JDACw/oD}    
&&19.84±0.53&0.1020±0.0062&0.7496±0.0236&0.9054±0.0126&28.94±0.54&0.0358±0.0022&0.6628±0.0288&0.9062±0.0114\\
{JDAC~(Ours)}    
& &25.51±1.34&0.0537±0.0089&0.8430±0.0245&0.9699±0.0128&\textbf{30.23±0.73}&\textbf{0.0309±0.0026}&\textbf{0.7571±0.0334}&0.9501±0.0142\\
\midrule

{JDACw/oA}    
&\multirow{3}{*}{$0.150$} &\textbf{25.50±1.28}&\textbf{0.0537±0.0087}&\textbf{0.8798±0.0296}&\textbf{0.9717±0.0131} &29.08±0.72&0.0353±0.0030&\textbf{0.7449±0.0352}&\textbf{0.9470±0.0152}\\
{JDACw/oD}    
&&18.92±0.49&0.1134±0.0063&0.7351±0.0236&0.8973±0.0124 &28.52±0.50&0.0376±0.0021&0.6455±0.0288&0.9024±0.0115\\
{JDAC~(Ours)}    
& &24.27±1.24&0.0618±0.0093&0.8174±0.0234&0.9613±0.0133 &\textbf{29.69±0.70}&\textbf{0.0329±0.0026}&0.7356±0.0321&0.9417±0.0143\\
\bottomrule
\end{tabular*}
\end{table*}
%%%%%%%%%%%%%%%%%%%%%%%%%%%%%%%%%%%%%%%%%%%%%%%%%%%%%%%%%%%%%%%%%%%%%%%%%%%%%%%%%%%%%%%%%%%%%%%%%%%%%%%%%%%%%%%%%%%%%%%%%%%%%
\subsection{Influence of Noise Level}
To investigate the adaptability of JDAC to different noise levels, we applied the JDAC and its two variants to the test set of MR-ART with manually added Gaussian noise of seven levels (\ie, noise std $\in [0.000, 0.025, \cdots, 0.150]$).
The results are reported in Table~\ref{ablation_noise_level}. 
\emph{On one hand}, as shown in the left part of Table~\ref{ablation_noise_level} for denoising, we can see that JDAC produces better results compared with its two counterparts when noise std is less than $<0.075$, while JDACw/oA yields the best results with noise std $\geq 0.100$. 
This implies that JDACw/oA contributes more when noise is not severe. 
\emph{On the other hand}, as shown in the right part of Table~\ref{ablation_noise_level} for motion artifact correction, JDAC achieves the overall best performance in terms of four metrics, while the JDACw/oD model cannot produce results with severe noise (\eg, noise std $=0.150$). 
These results suggest that our JDAC shows good adaptability to different noise levels, especially in the task of motion artifact correction.

%%%%%% - New Subsection - %%%%%%
\subsection{Influence of Motion Artifact Severity}
%%%%%%%%%%%%%%%%%%%%%%%%%%%%%%%%%%%%%%%%%%%%%%%%%%%%%%%%%%%%%%%%%%%%%%%%%%%%%%%%%%%%%%%%%%%%%%%%%%%%%%%%%%%%%%%%%%%%%%%%%%%%%
\begin{table}[!t]
\setlength{\belowdisplayskip}{-4pt}
\setlength{\abovedisplayskip}{0pt}
\setlength{\abovecaptionskip}{0pt}
\setlength{\belowcaptionskip}{0pt}
\scriptsize
%\tiny
\centering
\renewcommand{\arraystretch}{0.8}
\caption{Results of JDAC with different levels of motion artifacts (\ie, Gibbs artifact with a hyperparameter $\alpha$) severity in denoising and motion artifact correction task on MR-ART dataset. }
\label{ablation_motion_level}
\setlength{\tabcolsep}{2pt}
\begin{tabular}{l|c|cccc}
\toprule
\multirow{2}{*}{Method} &\multirow{2}{*}{Gibbs
}  &\multicolumn{4}{c}{Metrics on Gradient Map for Anti-Artifact Task}\\
\cmidrule(lr){3-6} 
&Artifacts &PSNR (dB) &RMSE  &SSIM &MS-SSIM\\
\midrule
{JDACw/oA}    
& \multirow{3}{*}{$\alpha=0.50$}  &42.21±0.84&0.0078±0.0007&0.9643±0.0063&0.9961±0.0009\\
{JDACw/oD}    
& &40.98±1.46&0.0091±0.0015&0.9318±0.0190&0.9656±0.0105\\
{JDAC~(Ours)}    
&&\textbf{44.67±1.63}&\textbf{0.0059±0.0011}&\textbf{0.9778±0.0044}&\textbf{0.9971±0.0009}\\
\midrule
{JDACw/oA}    
& \multirow{3}{*}{$\alpha=0.55$} %0.55 
&41.91±0.97&0.0081±0.0009&0.9624±0.0060&0.9956±0.0009\\
{JDACw/oD}    
& &41.19±1.42&0.0088±0.0015&0.9369±0.0169&0.9696±0.0100\\
{JDAC~(Ours)}    
&&\textbf{43.28±1.55}&\textbf{0.0070±0.0012}&\textbf{0.9703±0.0056}&\textbf{0.9959±0.0013}\\
\midrule

{JDACw/oA}    
& \multirow{3}{*}{$\alpha=0.60$}  &41.23±1.25&0.0088±0.0012&0.9554±0.0069&0.9940±0.0014\\
{JDACw/oD}    
&&41.36±1.34&0.0087±0.0015&0.9411±0.0144&0.9731±0.0092\\
{JDAC~(Ours)}    
&&\textbf{42.42±1.43}&\textbf{0.0077±0.0012}&\textbf{0.9642±0.0060}&\textbf{0.9948±0.0011}\\
\midrule

{JDACw/oA}    
& \multirow{3}{*}{$\alpha=0.65$} &40.82±1.33&0.0092±0.0013&0.9514±0.0078&0.9927±0.0017\\
{JDACw/oD}    
& &41.45±1.34&0.0086±0.0014&0.9436±0.0145&0.9762±0.0094\\
{JDAC~(Ours)}    
&&\textbf{41.69±1.44}&\textbf{0.0083±0.0013}&\textbf{0.9581±0.0071}&\textbf{0.9935±0.0014}\\
\midrule

{JDACw/oA}    
& \multirow{3}{*}{$\alpha=0.70$}%0.70
&40.26±1.33&0.0098±0.0014&0.9450±0.0080&0.9906±0.0018\\
{JDACw/oD}    
& &\textbf{41.31±1.23}&\textbf{0.0087±0.0013}&\textbf{0.9441±0.0132}&{0.9782±0.0088}\\
{JDAC~(Ours)}    
&&40.88±1.33&0.0091±0.0014&0.9507±0.0077&\textbf{0.9916±0.0017}\\

\midrule
{JDACw/oA}    
& \multirow{3}{*}{$\alpha=0.75$}%0.75
&39.55±1.37&0.0107±0.0016&0.9354±0.0096&0.9870±0.0022\\
{JDACw/oD}    
& &\textbf{40.79±1.02}&\textbf{0.0092±0.0010}&\textbf{0.9433±0.0083}&0.9795±0.0045\\
{JDAC~(Ours)}    
&&39.79±0.99&0.0103±0.0012&0.9397±0.0072&\textbf{0.9883±0.0018}\\
\midrule

{JDACw/oA}    
& \multirow{3}{*}{$\alpha=0.80$}  &38.60±1.23&0.0119±0.0016&0.9191±0.0112&0.9806±0.0029\\
{JDACw/oD}    
& &\textbf{39.68±0.75}&\textbf{0.0104±0.0009}&\textbf{0.9335±0.0082}&0.9738±0.0044\\
{JDAC~(Ours)}    
&&38.72±0.74&0.0116±0.0010&0.9250±0.0077&\textbf{0.9832±0.0024}\\
\bottomrule
\end{tabular}
\end{table}
%%%%%%%%%%%%%%%%%%%%%%%%%%%%%%%%%%%%%%%%%%%%%%%%%%%%%%%%%%%%%%%%%%%%%%%%%%%%%%%%%%%%%%%%%%%%%%%%%%%%%%%%%%%%%%%%%%%%%%%%%%%%%

To study the influence of motion artifact severity, we applied the JDAC and its two variants to the motion-free MRIs without noise in the test set of MR-ART. 
We generate Gibbs artifacts with different severity to each motion-free MRI using the GibbsNoise~\citep{morelli2011image} function in MONAI\footnote{https://docs.monai.io/en/stable/transforms.html\#gibbsnoise}, where the hyperparameter $\alpha$ is chosen from the range $[0.50, 0.60, \cdots, 0.80]$). 
The results of JDAC and its variants are reported in Table~\ref{ablation_motion_level}. 
From Table~\ref{ablation_motion_level}, JDAC yields the best results when the artifact level is not severe (\ie, $\alpha \leq 0.65$), and achieves comparable results with the second-best JDACw/oD method when $\alpha \geq 0.70$. 
It is suggested that our JDAC is not very sensitive to different levels of artifact severity. 
\emph{In addition}, it can be seen from the right part of Table~\ref{comparison_MRART} and Table~\ref{ablation_motion_level} that the overall performance of JDAC is better when evaluated on MRIs with simulated artifacts. 
This implies that it is be more challenging to model and eliminate real motion artifacts as opposed to simulated ones. 

%%%%%% - New Subsection - %%%%%%
%%%%%%%%%%%%%%%%%%%%%%%%%%%%%%%%%%%%%%%%%%%%%%%
\begin{figure}[!t]
\setlength{\belowdisplayskip}{0pt}
\setlength{\abovedisplayskip}{0pt}
\setlength{\abovecaptionskip}{0pt}
\setlength{\belowcaptionskip}{0pt}
\centering
\includegraphics[width=0.48\textwidth]{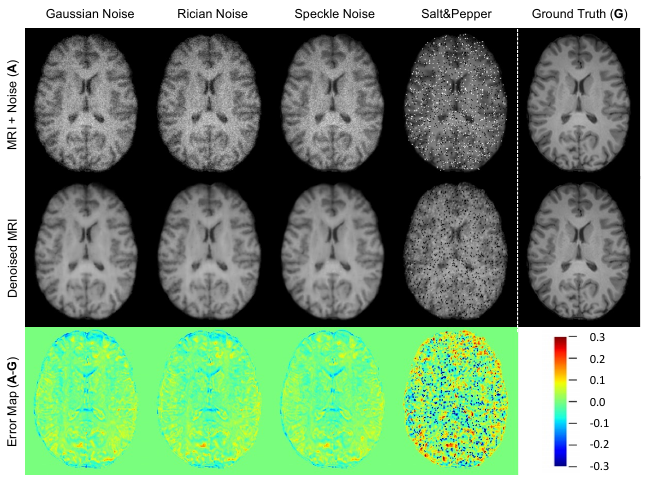}
\caption{Denoising results of JDAC on four types of simulated noise in MRI (\ie, Gaussian, Rician, Speckle, and Salt\&Pepper noise).}
\label{fig_vasualize_NoiseType}
\end{figure}
%%%%%%%%%%%%%%%%%%%%%%%%%%%%%%%%%%%%%%%%%%%%%%%
\subsection{Influence of Noise Type}
To study the influence of different noise types, we further conduct an experiment by adding four kinds of simulated noise to each MRI, including (1) Gaussian noise~\citep{aja2016statistical}, (2) Rician noise~\citep{gudbjartsson1995rician}, (3) Speckle noise~\citep{pizurica2006review}, and (4) Salt\&Pepper noise~\citep{ebrahimnejad2021adaptive}.   
The visualization of denoising results generated by JDAC is shown in Fig.~\ref{fig_vasualize_NoiseType}.  
This figure suggests that JDAC performs better when facing the first three types of noise with relatively similar distributions (\ie, Gaussian noise, Rician noise, and Speckle noise), but cannot well handle Salt\&Pepper noise that exhibits significantly different distribution with Gaussian noise. 
The possible reason is that JDAC is trained on images with Gaussian noise, and thus, cannot generalize well to MRIs corrupted by Salt\&Pepper noise with binary values.  
It is interesting to re-train or fine-tune our model on data with diverse types of noise, further improving its generalizability to different problems.

\subsection{Influence of Motion Artifact Type}
\begin{figure}[!t]
\setlength{\belowdisplayskip}{0pt}
\setlength{\abovedisplayskip}{0pt}
\setlength{\abovecaptionskip}{0pt}
\setlength{\belowcaptionskip}{0pt}
\centering
\includegraphics[width=0.48\textwidth]{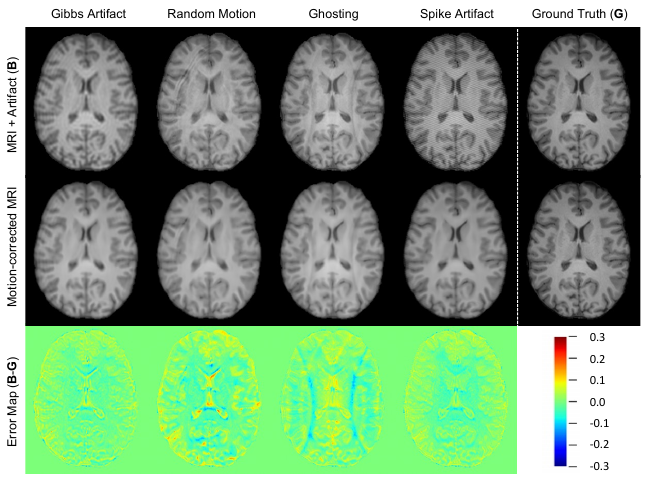}
\caption{Anti-artifact results of JDAC on four types of simulated artifacts in MRI (\ie, Gibbs artifact, Random motion, Ghosting, and Spike artifact). }
\label{fig_vasualize_ArtifactsType}
\end{figure}

To study the influence of different types of motion artifacts, we perform experiments by applying JDAC to MRIs with four kinds of common artifacts, including (1) Gibbs artifact~\citep{morelli2011image}, (2) Random motion ~\citep{shaw2019mri}, (3) Ghosting~\citep{discreteghosts}, and (4) Spike artifact~\citep{graves2013body}). 
All these artifacts are simulated using tools in MONAI~\citep{cardoso2022monai}\footnote{https://docs.monai.io/en/stable/transforms.html} and TorchIO~\citep{perez2021torchio}\footnote{https://torchio.readthedocs.io/}. 
For a fair comparison, we set some hyperparameters to generate artifacts of middle severity: 
(1) $\alpha = 0.70$ for Gibbs artifact;
(2) rotation degree within [5, 8], translations along each axis within [3, 5]$mm$, and the number of transforms of 4 for random motion; 
(3) number of ghosts within [4, 10] and intensity strength within [0.5, 1] for random Ghosting; and  
(4) number of spikes of 1 and intensity strength of 0.5 for random Spike artifact. 
In Fig.~\ref{fig_vasualize_ArtifactsType}, we visualize a typical MR image corrupted by simulated artifacts, as well as the anti-artifact results generated by JDAD. 
Observing the lower two rows of this figure, it is evident that JDAC effectively addresses Gibbs and Spike artifacts, in contrast to Random Motion and Ghosting which introduce more uneven distortions to the local brain anatomy.

%%%%%% - New Subsection - %%%%%%
%%%%%%%%%%%%%%%%%%%%%%%%%%%%%%%%%%%%%%%%%%%%%%%
%%%%%%%%%%%%%%%%%%%%%%%%%%%%%%%%%%%%%%%%%%%%%%%

\subsection{Computation and Time Cost Analysis}
We further calculate the computation costs of our JDAC and all competing methods using the Flopth toolbox\footnote{https://github.com/vra/flopth}, and report the average time spent processing each test MRI in Table~\ref{tab_cost}. 
Owing to GPU memory constraints, the time cost is recorded based on the same CPU platform (Intel(R) Core (TM) i7-8700K CPU @ 3.70GHz). 
It can be seen from Table~\ref{tab_cost} that the parameter scales and FLOPs of all the five 3D models are usually higher than that of 2D-based methods (\ie, DRN-DCMB and SUNet), while 2D-based methods are more time-consuming during inference because they need to process 3D data slice-by-slice.
The FONDUE and BM4D methods have the highest test time costs since they require data sampling repeatedly during processing. 
\emph{On the other hand}, Table~\ref{tab_cost} indicates that although JDAC necessitates longer training time compared to UNet3D and nnUNet, it manages to attain a comparable test time cost (\ie, 5.80 seconds) to that of the two 3D models (\ie, 3.30 seconds for UNet3D and 3.02 seconds for nnUNet). 
The possible reason is that while JDAC, as an iterative framework, generally demands more training time, it typically needs only 1-2 iterations during inference, thanks to the proposed early stopping strategy.

\subsection{Limitations and Future Work}
Several issues need to be considered in the future. 
\emph{First}, the proposed denoising and motion artifact correcting models tend to smooth out some brain structure details.
Thus, we will consider other state-of-the-art methods to further preserve the brain anatomy structures in MR images during iterations in the future. 
\emph{Second}, the severity of motion artifacts in MRIs varies from subject to subject.
As the severity of artifacts increases, there is a notable decline in the performance of all anti-artifact models. 
Intuitively, detecting motion artifacts and assessing their severity could enhance the effectiveness of the artifact reduction model.
Accordingly, we will incorporate advanced artifact detection/assessment methods~\citep{jimeno2022artifactid,haskell2019network} in JDAC to adaptively reduce motion artifacts based on the severity of motion artifacts. 
\emph{In addition}, we train and validate the JDAC only on T1-weighted brain MRIs in experiments of the current work. 
As an interesting future work, we will more extensively evaluate our method on 3D medical images of other modalities.

%%%%%%%%%%%%%%%%%%%%%%%%%%%%%%%%%%%%%%%%%%%%%%%
\begin{table}[!t]
\centering
\setlength{\abovecaptionskip}{0pt}
\setlength{\belowcaptionskip}{0pt}
\setlength{\abovedisplayskip}{0pt}
\setlength{\belowdisplayskip}{0pt}
%\small
\footnotesize
\caption{Time and computation cost. For JDAC, $a+b$ denotes the numbers for the denoising model and the anti-artifact model. M: Million; GMac: Giga multiply-accumulate operations; H: Hour; S: Second.%; -: not applicable.
}
%\footnotesize
\scriptsize
\renewcommand{\arraystretch}{0.8}
\setlength{\tabcolsep}{2pt} 
\begin{tabular}{l|cccc}
%\begin{tabular}{l|p{1.7cm}p{1.5cm}p{1.2cm}p{1cm}p{1cm}}
\toprule
Method & {$\#$Parameters (M)} & {FLOPs (GMac)}    & {Training Time (H)}  & {Test Time (S)} \\
\midrule
{DRN-DCMB} &0.11 &1.86   &6.67  &27.03 \\
{SUNet}  &4.08 &25.73  &13.43 &81.74 \\
{BM4D}     &-    &-      &-     &109.63\\
{UNet3D}   &3.33 &130.86 &4.99  &3.30  \\
{nnUNet}   &3.75 &132.47 &7.44  &3.02  \\
{FONDUE}   &2.16 &69.10  &-     &227.33\\
\textbf{JDAC~(Ours)}   & 2.92+2.92 & 149.41+149.34 & 13.19+9.32 &5.80 $\sim$ 15.50   \\
\bottomrule
\end{tabular}
\label{tab_cost}
\end{table}
%%%%%%%%%%%%%%%%%%%%%%%%%%%%%%%%%%%%%%%%%%%%%%%

%%%%%%%%%%%%%%%%%%%%%%%%%%%% new section   new section  new section %%%%%%%%%%%%%%%%%%%%%%%%%%%%%%%%%%%
\section{Conclusion}
\label{S6}
This paper introduces a joint image denoising and motion artifact correction (JDAC) framework to iteratively process noisy brain MRIs with motion artifacts. 
The JDAC consists of an adaptive denoising model and a motion artifact correction model, where the noise level of each input MRI is explicitly estimated based on the intensity variance of its gradient map. 
We validate JDAC on two public datasets and a clinical study with motion-affected MRIs, with experimental results suggesting its effectiveness in both tasks of denoising and motion artifact correction, compared with several state-of-the-art methods. 

\section*{Acknowledgments}
Part of the data used in this paper was obtained from the Alzheimer's Disease Neuroimaging Initiative (ADNI). 
The ADNI investigators contributed to the design and implementation of ADNI and provided data but did not participate in the analysis or writing of this article. 
A full list of ADNI investigators can be found online ({\small{https://adni.loni.usc.edu/wp-content/uploads/how\_to\_apply/ADNI\_Acknowledgement\_List.pdf}}).

\vspace{-4pt}
\small
\bibliographystyle{model2-names.bst}
\biboptions{authoryear}
\bibliography{refs}

\end{document}